\documentclass[%
 reprint,
 amsmath,
 amssymb,
 aps,
 prb,
]{revtex4-2}

\usepackage{graphicx}
\usepackage{dcolumn}
\usepackage{bm}
\usepackage{xcolor}
\usepackage{braket}

\begin{document}

\preprint{APS/123-QED}

\title{Role of the edges in a quasicrystalline Haldane model}

\author{Simone Traverso$^{(1)}$}
%
\author{Maura Sassetti$^{(1,2)}$}
\author{Niccol\`o Traverso Ziani$^{(1,2)}$}
\affiliation{1 Dipartimento di Fisica, Universit\`a degli studi di Genova, Via Dodecaneso 33, 16146 Genova GE, Italy}
\affiliation{2 CNR-SPIN, Via Dodecaneso 33, 16146 Genova GE, Italy}


\date{\today}

\begin{abstract}
We study the role of the edges in determining the features of the topological phase in a quasicrystalline higher order topological insulator.
We consider a specific model consisting of two stacked Haldane models with opposite Chern number and a $30^\circ$ twist, whose structure is crystallographically equivalent to that of the graphene quasicrystal. We find that the gap-opening in the low energy spectrum of the higher order topological insulator occurs at different energies when different kinds of edges are considered. Crucially, bearded bonds appear to be necessary for the gap to appear close to the charge neutrality point. In the more realistic case of zigzag edges, the gap opens symmetrically in the electron and hole sectors, away from zero energy.
We explain our findings by inspecting the edge-bands of the decoupled bilayer, in the approximation of quasi-periodicity.
\end{abstract}

\maketitle


\section{\label{sec:intro}Introduction}
The usual distinction between conductors and insulators has been surpassed with the discovery of topological insulators \cite{RevModPhys.82.3045, PhysRevLett.98.106803, doi:10.1126/science.1133734, Zhang:2009zzf}, in which a bulk band gap coexists with protected conducting states localized at the boundary of the system. In ordinary topological insulators, the metallic edge states propagate in one dimension less than the dimensionality of the bulk, in accordance to the principle of \emph{bulk-boundary correspondance} \cite{bernevig2013topological}. Moreover, typically, elastic backscattering within the topological modes is forbidden, for geometric or symmetry reasons~\cite{PhysRevLett.95.146802, PhysRevLett.95.226801}. This fact makes topological insulators interesting for quantum technological applications~\cite{He2019, Hsieh2009, Mellnik2014}.

Recently, new topological phases have been discovered, characterized by protected metallic states that live in a dimension $D-d$, with $d>1$ ($D$ being the dimension of the bulk). Systems that manifest such feature are called \emph{higher order topological insulators} (HOTI), and $d$ is referred to as the order of the topological phase \cite{doi:10.1126/sciadv.aat0346, doi:10.1126/science.aah6442, PhysRevB.96.245115, PhysRevLett.119.246401, doi:10.1126/sciadv.aat2374}. To date, HOTI phases have been experimentally achieved in 3D materials \cite{noguchiHoti, schindlerHOTI}, while a realization on 2D materials is still missing.

In the search for platforms able to host the HOTI phase, quasicrystals~\cite{doi:10.1126/science.1170827} were rapidly found to be promising candidates. Indeed, although quasicrystals lack translational invariance, they still possess long range order. Such an order may express itself in the form of a rotational symmetry not achievable in crystals, thus opening new opportunities for symmetry protection. In the last couple of years, a few models of HOTI on quasicrystals have been proposed, mainly for $D=d=2$ \cite{PhysRevLett.124.036803, PhysRevLett.123.196401, PhysRevResearch.2.033071}.

Among these models, the one for a (spinless) second order topological insulator (SOTI) proposed by S. Spurrier and N. Cooper in \cite{PhysRevResearch.2.033071} is particularly relevant, being its potential experimental realization almost at reach~\cite{maciver,doi:10.1126/science.aar8412,claessen}. Indeed, the lattice on which the model develops is crystallographically equivalent to the one characterizing the \emph{graphene quasicrystal} (GQ), a Van der Waals structure belonging to the class of \emph{twisted bilayer graphene} (TBG) systems. GQ is composed of two stacked graphene layers with a relative rotation angle of $30^\circ$ and it has been experimentally realized in 2018~\cite{doi:10.1126/science.aar8412}. In the same occasion, it has been proven to possess 12-fold rotational symmetry ($C_{12}$) together with a peculiar electronic structure \cite{doi:10.1126/science.aar8412} consisting of twelve Dirac cones. Moreover, recent advances in the experimental techniques have made it possible to directly inspect the edges of 2D graphene-like lattices in the topological regime~\cite{claessen}, and hence the direct detection of the HOTI phase- that requires the resolution of the zero-dimensional states -would also be possible. 
A point which is potentially relevant but has not been discussed in the proposal by Spurrier and Cooper is the effect that the shape of the edges has on the opening of the gaps protecting the topological modes, and hence their experimental accessibility. Indeed, the influence of the edges in finite size graphene samples is well known \cite{RevModPhys.81.109, PhysRevLett.119.076401}.

In this paper, we focus on this aspect, studying how the shape of the edges modifies the features of the topological phase. Since we require that the sample edges are cropped in a way which is compatible with the bulk symmetry of the GQ-like lattice, the edges of the single honeycomb layers can be cut in three possible ways: armchair, zigzag and bearded~\cite{PhysRevLett.119.076401}. Consequently, in the bilayer configuration, the possible couples of edges are armchair-zigzag and armchair-bearded (the case inspected in \cite{PhysRevResearch.2.033071}). By means of a tight binding model, we find that the gaps hosting the second order topological modes open at different energies in the two cases. In particular, we show that the gap opens at the charge neutrality point in the armchair-bearded case, while it moves away from zero energy in the armchair-zigzag scenario. Finally, we give an explanation of our results by inspecting the bands of the uncoupled bilayer model through the quasiperiodic approximation. {\color{black} Our results are significant for the experimental feasibility of the proposal in~\cite{PhysRevResearch.2.033071}. Moreover, their impact extends to experiments such as the one in~\cite{claessen}, where different edges of honeycomb lattices in the topological regime are in close proximity.}

The rest of the paper is structured as follows: In Sec.~\ref{sec:model} we discuss the properties of the lattice and we introduce the Hamiltonian of the model. In Sec.~\ref{sec:spect} we show the results obtained through numerical diagonalization for the different configurations of the edges. In Sec.~\ref{sec:interpretation} we interpret them on the basis of simple physical arguments. Finally, in Sec.~\ref{sec:conc} we present our conclusions.

\section{\label{sec:model} Model}

With the aim of studying the edge dependence of the model for a SOTI on quasicrystal proposed in \cite{PhysRevResearch.2.033071}, in this Section we first build the finite size quasicrystalline bilayer lattice, and then we present the tight-binding model that develops on it.

\subsection{\label{subsec:lattice} Lattice}
\begin{figure}[hbt]
    \centering
    \includegraphics[width=\columnwidth]{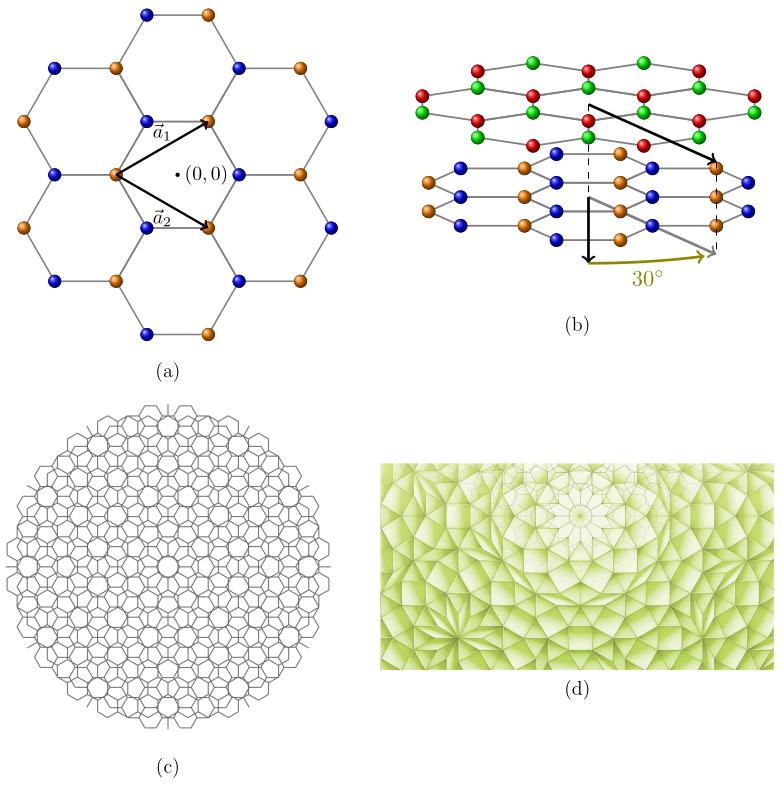}
    \caption{Panel (a): scheme of a single layer of honeycomb lattice. The origin of the coordinate axes is placed at the center of a hexagon. Panel (b): scheme of the twisted bilayer construction; starting from an AA stacking the upper plane is rotated of $30^\circ$ around an axis that passes through the center of a couple of superimposed hexagons. Panel (c): scheme of a sample of the quasicrystalline bilayer reported in Panel (b), with the upper layer projected on the lower one. Instead of the sites, the links between first neighbors on each plane are shown. Panel (d), adapted from~\cite{science_stampfli}: the Stampfli tiling superimposed to the quasicrystalline lattice; the $C_{12}$ symmetry of the tiling is evident.}
    \label{fig:lattice}
\end{figure}

In the process of building the lattice, we start from the single honeycomb layer. We choose as primitive vectors
\begin{equation}
    \vec{a}_1 = a\left(\dfrac{\sqrt{3}}{2},\frac12\right), \qquad \vec{a}_2 = a\left(\dfrac{\sqrt{3}}{2},-\frac12\right),
\end{equation}
with $a$ the next-nearest distance in the honeycomb lattice, and we set the origin of our axes at the center of one of the lattice hexagons. A scheme of the lattice cell is shown in Fig.~\ref{fig:lattice}(a). We observe \emph{en passant} that the honeycomb lattice alone already possess a bulk $C_6$ rotational symmetry. This means that the lattice is invariant for rotations of $60^\circ$, up to the exchange of the two sublattices of which it is made of.

We now proceed to add a second honeycomb layer on the top of the first one, in such a way that each site is perfectly superimposed to the corresponding one of the original layer. In multilayer graphene systems this particular configuration is referred to as AA stacking. Once this is done, we rotate the new layer of $30^\circ$ around an axis perpendicular to the two layers and containing the origin, as shown in Fig.~\ref{fig:lattice}(b). A representation of the total lattice can be found in Fig.~\ref{fig:lattice}(c).

The structure obtained with the procedure just described, in the limit $d_0\ll L$, with $d_0$ the interlayer separation and $L$ a scale of length of the lattice sample, is indeed a quasicrystal. In fact, as shown in Fig.~\ref{fig:lattice}(d), it can be mapped onto the Stampfli quasicrystal~\cite{schaad2021quasiperiodic}, an aperiodic tiling that has been proven to possess $C_{12}$ local symmetry.

Since our goal is to achieve a SOTI phase on the lattice just discussed, we now turn to the problem of picking up a finite size sample. While any finite sample of this lattice would possess bulk $C_{12}$ symmetry, in order to observe the second order topological phase described in \cite{PhysRevResearch.2.033071} a global rotational symmetry is needed~\cite{ncc}. This constraint forces us to choose a shape that is compatible with the $C_{12}$ bulk symmetry, meaning that it must have a $C_n$ rotational symmetry that respects $C_n=C_{12}^q$, for some positive integer $q$. Such a requirement can only be met by a few polygonal shapes: dodecagonal, hexagonal, square and triangular. However, as we mentioned above, the $C_3$ and $C_6$ global rotational symmetry, possessed by the hexagonal and triangular shape respectively, are already present in the single-layer honeycomb lattice. Therefore since they are not unique to the quasicrystalline lattice, we discard them, being of less interest in the current context. Finally, among the two shapes left available, in what follows we will take the square one for our samples, since this choice brings some computational advantages. Nevertheless it must be noted that the results we will discuss have their exact counterparts in the case of dodecagonal samples, {\color{black} as explicitly discussed in the Appendix. Moreover, our analytical interpretation gives a hint in favor of the fact that our results only depend on the shape of the edges, and not on the geometry}.

A further crucial point needs to be addressed, that is the one regarding the orientation of the cropping shape with respect to the lattice. We would like to orient the shape so that the edges of the bilayer lattice are not jagged after we have cropped them: this is because if all the sites on each edge of the single layers are lined up, a much better superposition can be achieved between the upper and lower edge. This will reveal to be crucial in order to achieve the SOTI phase.

\begin{figure}
    \centering
    \includegraphics[width=\columnwidth]{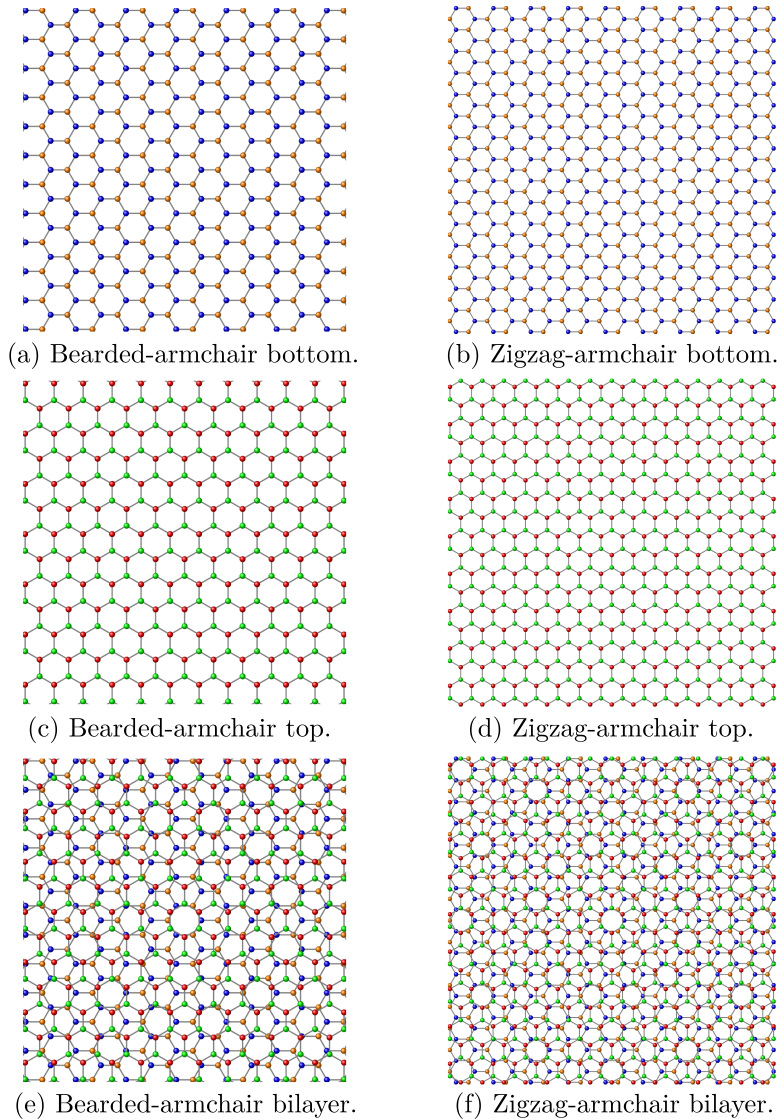}
    \caption{Here are shown two distinct samples of the quasicrystalline lattice, that exemplify the two possible combinations of edges described in the main text. In Panels (a) and (c) are reported respectively the bottom and top layer for the case of bearded-armchair edges, while their combination in the resulting quasicrystalline bilayer lattice is shown in Panel (e). Panels (b), (d) and (f) are the corresponding representations for a sample with zigzag-armchair edges. In both cases the upper and lower layer are just the same, up to a rotation of $90^\circ$. The bilayer structure presents bulk $C_{12}$ local symmetry and, due to the square shape chosen, global $C_4$ symmetry.}
    \label{fig:bilayer}
\end{figure}

It turns out that the only ``non-jagged'' edges that can be obtained from a honeycomb lattice are of armchair, zigzag and bearded type. Moreover, one finds that, due to the geometry of the honeycomb lattice, a square shape can be oriented in such a way that \emph{all} the edges are of armchair, bearded or zigzag type, without any of them being jagged \footnote{It is worth pointing out that this fact is true for a dodecagonal shape as well. This makes so that, as stated above, the following analysis can be replicated for samples with dodecagonal geometry.}. In Fig.~\ref{fig:bilayer}, two square samples are shown, together with the corresponding separate layers. The two configurations differ for the type of their edges: It can be seen that on the square sample of monolayer, an armchair edge is always followed by one of zigzag or bearded type; which one of these two depends on the length of the cropping square edge. More importantly, for what it concerns the bilayer we find that an armchair edge in one layer is always coupled with one of zigzag or bearded type in the other layer, and viceversa. No other configuration can be obtained for non-jagged edges.

In the following we will study how the higher order topological phase achievable on this quasicrystalline lattice is influenced by the edge configuration adopted. Anyway, before we can come to that we must still introduce the model Hamiltonian. This will be the task of the next Subsection.

\subsection{\label{subsec:ham} Model Hamiltonian}
The HOTI model we analyze \cite{PhysRevResearch.2.033071} is encoded in the Hamiltonian
\begin{equation}
	H= t \sum_{\langle ij\rangle}c_i^\dagger\tau_0c_j +\lambda_H\sum_{\langle\langle ij\rangle \rangle}i\nu_{ij}c_i^\dagger \tau_z c_j +\lambda_{\perp}\sum_{ij}t^\perp_{ij}c_i^\dagger\tau_x c_j.
	\label{eq:spurrier_ham}
\end{equation}
Here, $c_i=(c_i^{t},c_i^{b})^T$ is a two component object, composed of the spinless Fermionic operators $c_i^{t/b}$ destroying a fermion on the $i$-th lattice site of the top/bottom layer. Moreover, the Pauli matrices $\tau_i$ act in the space of layer isospin. The fermions described by $H$ undergo in-plane nearest neighbors ($\langle ij\rangle$) hopping with amplitude $t$, and next to nearest neighbor ($\langle\langle ij\rangle \rangle$) hopping of amplitude $\lambda_H$, with an associated complex phase $i\nu_{ij}$ ($\nu_{ij}=\pm 1$) given by the usual Haldane coupling~\cite{PhysRevLett.61.2015}. The explicit form of $\nu_{ij}$ is given for example in~\cite{bernevig2013topological}. Moreover, an inter-layer hopping is present, parameterized by $t^\perp_{ij}$. Its explicit form is derived from tight-binding models for real TBG samples~\cite{PhysRevB.85.195458, PhysRevB.87.205404}
\begin{equation}
    t_{ij}^\perp = t^\perp \exp \left(-\frac{|\vec{d}_{ij}|-d_{0}}{\delta}\right),
    \label{eq:t_perp}
\end{equation}
where $\vec{d}_{ij}$ is the vector that connects the $i$ and $j$ sites on the two planes; $\delta = 0.184 a$ is the decay length of the transfer integral (expressed in function of the in-plane next-nearest neighbor distance $a$); $d_0=1.362 a$ is the interlayer distance and $t^\perp=-0.178 t$ is a coupling coefficient~\footnote{$t_\perp$ is chosen so that for $\lambda_\perp=1$ the coupling for interlayer vertical hoppings has the same numerical value as in \cite{PhysRevB.85.195458, PhysRevB.87.205404}}. In Eq.~(\ref{eq:spurrier_ham}) the parameter $\lambda_\perp$ is added in order to tune the interlayer coupling intensity.

As already underlined in \cite{PhysRevResearch.2.033071} the Hamiltonian consists of two Haldane models~\cite{PhysRevLett.61.2015} with opposite Chern number (\emph{i.e.}~a Kane-Mele model~\cite{PhysRevLett.95.226801, PhysRevLett.95.146802} with the spin replaced by layer isospin) and with a $30^\circ$ twist, coupled together through the interlayer hopping. Without this last term, the low energy spectrum would be gapless, due to the presence of counterpropagating edge modes at the boundaries of the two planes, just as in the Kane-Mele model. Indeed, as will be shown in the next Section, the interlayer coupling gaps out the counter-propagating modes, making it possible for the SOTI phase to occur.

\section{\label{sec:spect} Spectral properties}
We will now present the results obtained by diagonalizing the Hamiltonian in Eq.~(\ref{eq:spurrier_ham}) for finite-size square samples with the two possible kinds of edges discussed above. The results reported in this Section have been obtained through a numerical tight-binding approach, using the package \emph{pybinding} for the construction and diagonalization of the system Hamiltonian~\cite{dean_moldovan_2020_4010216}. {\color{black} About the coupling parameters in Eq.~(\ref{eq:spurrier_ham}), we have set $\lambda_H=0.3t$ to maximize the Haldane gap while still avoiding band overlapping~\cite{PhysRevLett.61.2015}, and $\lambda_\perp=2$. It is worth noting that the results obtained would have been qualitatively the same for different values of $\lambda_\perp$. Indeed, to a certain extent, tuning $\lambda_\perp$ would only modify the amplitude of the gap opening between the edge modes \footnote{As far as $\lambda_H$ is concerned, the request is to be in the Chern insulator regime. In the absence of an on-site chemical potential, this condition is met, in the full Haldane model, for $\lambda_{H}/t \in ]0,1/3[$. However, a sizable $\lambda_H>0.1t$ is beneficial for making the interpretation clearer, since for smaller values the topological edge states in the zigzag and bearded edges become reminiscent of the non-dispersive ones that characterize the $\lambda_H=0$ regime. Moreover, for our sample sizes we have explicitly checked the stability of the results for $\lambda_\perp \in [1,3]$. Bigger sample sizes allow for a larger parameter range, since the hybridization of the corner modes is reduced by their increased spatial separation.}.} Moreover, in order to avoid unnecessary computational complexity, the interlayer coupling has been limited to sites whose relative distance $|\vec{d}|$ satisfies $|\vec{d}|-d_0<2a_{cc}$, with $a_{cc}=a/\sqrt{3}$ the in-plane nearest neighbor distance. Indeed, over this distance the effects of the interlayer interaction can be safely neglected.

\begin{figure}
    \centering
    \includegraphics[width=\columnwidth]{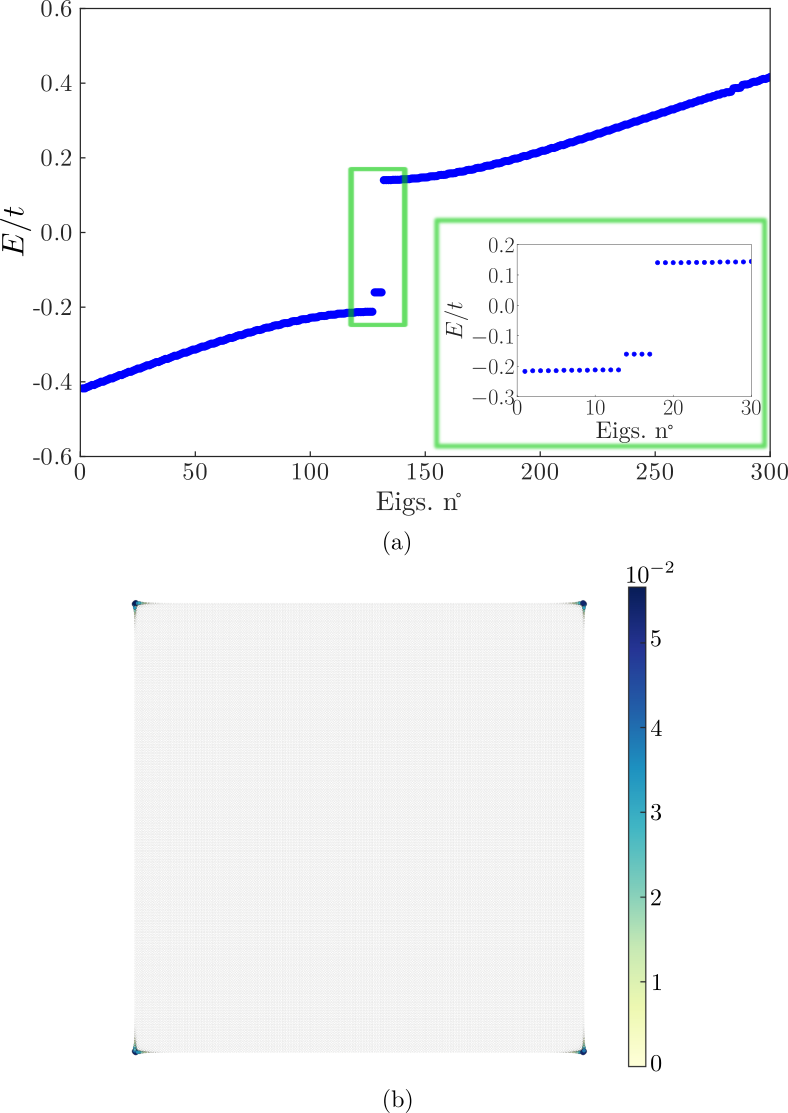}
    \caption{Panel (a): Low energy eigenvalues obtained by diagonalizing Hamiltonian (\ref{eq:spurrier_ham}) for a square sample with armchair-bearded edges of length $L=433 a_{cc}$, for $\lambda_\perp=2$ and $\lambda_H=0.3t$. The gap opens around $E\approx 0$ and hosts four degenerate eigenvalues, as can be seen in the inset. Panel (b): Density plot of the square modulus of an eigenstate associated to one of the degenerate in-gap eigenvalue. Choosing a different bound state would produce a similar result.}
    \label{fig:spect_dang}
\end{figure}

\begin{figure}
    \centering
    \includegraphics[width=\columnwidth]{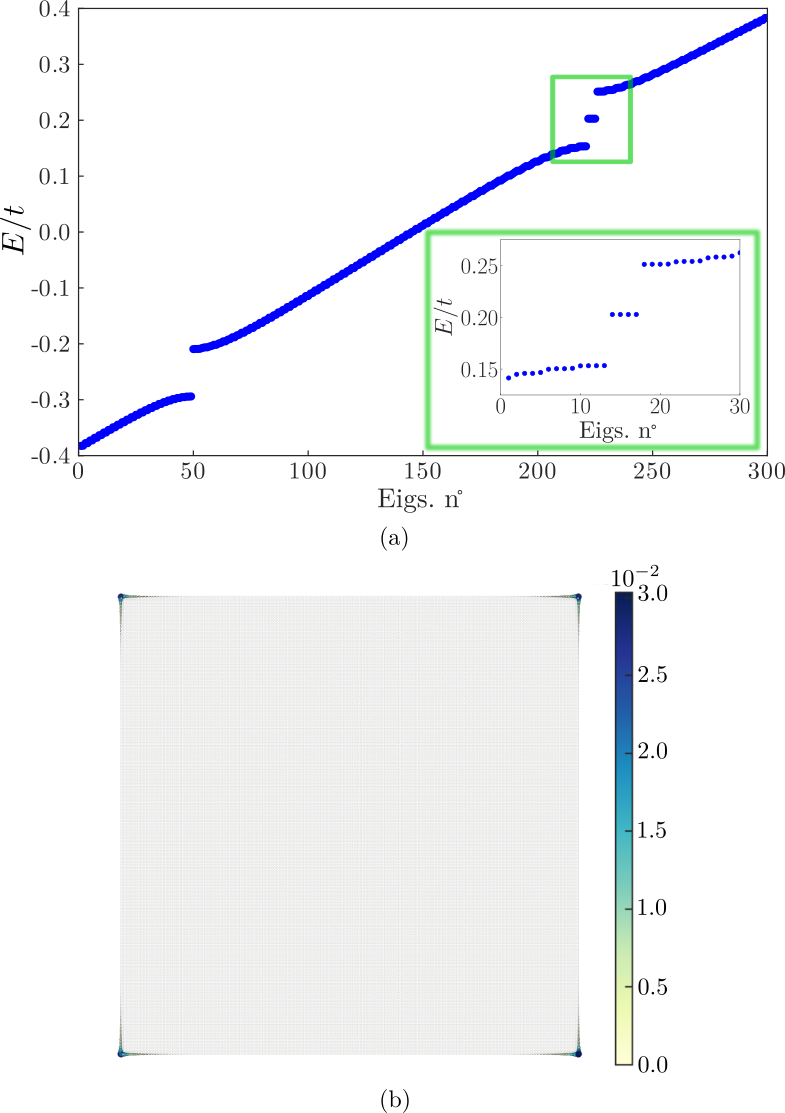}
    \caption{Panel (a): Low energy eigenvalues obtained by diagonalizing Hamiltonian (\ref{eq:spurrier_ham}) for a square sample with armchair-zigzag edges of length $L=485 a_{cc}$, for $\lambda_\perp=2$ and $\lambda_H=0.3t$. Two gaps open around $E/t\approx \pm 0.2$ and the one at higher energy hosts four degenerate eigenvalues, as can be seen in the inset. Panel (b): Density plot of the square modulus of an eigenstate associated to one of the degenerate in-gap eigenvalue. Choosing a different bound state would produce a similar result.}
    \label{fig:spect_zig}
\end{figure}

In Fig.~\ref{fig:spect_dang}(a) the spectrum obtained for a square sample whose edges are of armchair-bearded type is reported. A gap opens around $E/t=0$ due to the interlayer interaction, that gaps out the counter-propagating  Haldane modes on the edges of the two planes. Inside the gap four degenerate eigenvalues are found, whose position in energy perfectly matches the one expected by applying the low energy theory developed in \cite{PhysRevResearch.2.033071} in the case of global $C_4$ symmetry. These 4-fold degenerate eigenvalues correspond to 0-dimensional (0D) corner modes, localized on the vertices of the square sample (see Fig.~\ref{fig:spect_dang}(b)). This is the most favorable case for the realization of the SOTI in the system.

In Fig.~\ref{fig:spect_zig}(a) the spectrum obtained for a square sample with armchair-zigzag edges is reported. The difference with respect to the previous scenario (Fig.~\ref{fig:spect_dang}) is remarkable: Indeed, we see that with this choice of the edges two gaps- smaller than the one found in the armchair-bearded case -open in the spectrum, at $E/t\approx\pm 0.2$ respectively~\footnote{More precisely, the gaps would open around $E/t\approx +0.2$ and $E/t\approx -0.25$. The tiny asymmetry in the gap openings is due to the interlayer hopping term which, once normalized with respect to $t$, has a negative coefficient ($t^\perp=-0.178 t$). This makes so that the whole spectrum is slightly shifted down. For simplicity, in what follows we will always denote the energy of the two gaps just as $E/t\approx \pm 0.2$}. In the one at higher energy four degenerate eigenvalues can still be found. The corresponding eigenstates are localized on the vertices of the sample, as before (Fig.~\ref{fig:spect_zig}(b)).

Increasing the sample sizes one finds that the position of the degenerate eigenvalues inside the gaps, measured with respect to the midgap energy ($E_{\text{mg}}$) and normalized to half the gap width ($m$), converges to a specific value, that is $\frac{E-E_{\text{mg}}}{m}=-\frac{1}{\sqrt{2}}$ for the case of the armchair-bearded edges- as reported in \cite{PhysRevResearch.2.033071} -and $\frac{E-E_{\text{mg}}}{m}=0$ for the case of armchair-zigzag edges.

\section{\label{sec:interpretation} Interpretation}
In this Section, we will give a physical argument to explain the reason why the gap opens around different energies depending on the shape of the edges. The reasoning will be divided in three separate steps:
First, we will analyze the bands of the Haldane model on a single strip, that is a ribbon of honeycomb lattice, infinite in one direction and finite and uncompactified {\color{black}(no periodic boundary conditions)} in the other. We will discuss separately the spectral properties of the model in the case of armchair, zigzag and bearded edges along the uncompactified terminations.
Then, from the monolayer strip bands, the bands of the (uncoupled) bilayer edge will be obtained in the quasiperiodic approximation. From these, the possible positions where the edge gaps may open will be deduced. Finally, by adopting a low energy effective theory, the hierarchy of the possible gaps will be discussed. This will lead us to predict the position of each possible gap and the order in which they would open if one were to indefinitely increase the sample size. 

\subsection{Haldane model on a strip}
As anticipated, we will now consider the Haldane model on a strip for the three possible kinds of edges discussed in Subsec.~\ref{subsec:lattice} and proceed to diagonalize the corresponding Hamiltonian with periodic boundary conditions, in order to obtain the energy bands for each different edge type.

\begin{figure}
    \centering
    \includegraphics[width=\columnwidth]{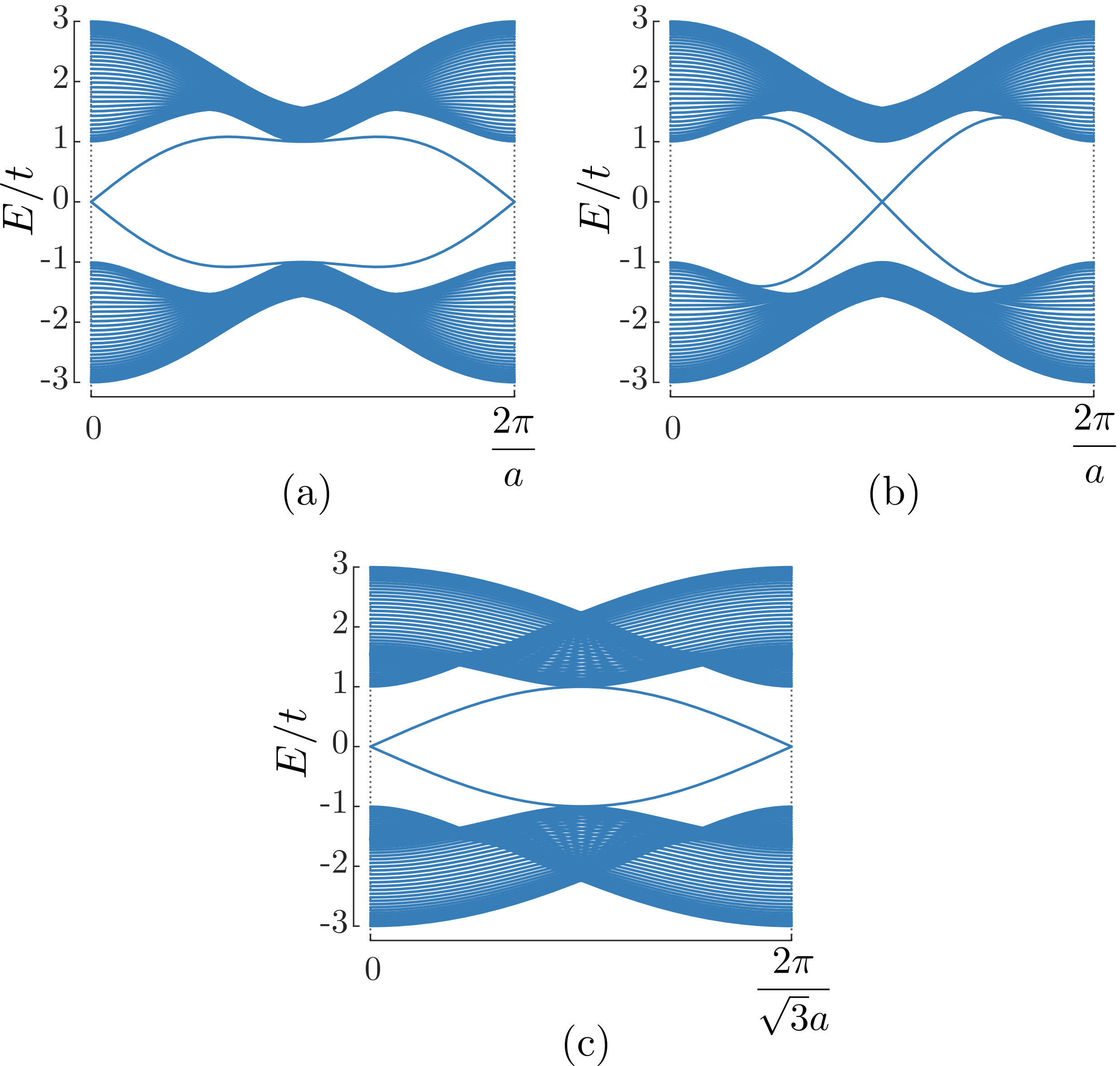}
    \caption{Bands for the Haldane model on a strip geometry, with $\lambda_H= 0.3t$. In Panel (a), (b) and (c) are reported the bands for a strip with bearded, zigzag and armchair edges respectively.}
    \label{fig:bands}
\end{figure}

The Hamiltonian for the Haldane model is \cite{PhysRevLett.61.2015}
\begin{equation}
    H= t \sum_{\langle ij\rangle}\tilde{c}_i^\dagger \tilde{c}_j +\lambda_H\sum_{\langle\langle ij\rangle \rangle}i\nu_{ij}\tilde{c}_i^\dagger \tilde{c}_j,
    \label{eq:hald}
\end{equation}
$\tilde{c}_i^\dagger$ being the creation operator for a spinless fermion on the honeycomb lattice, and $i\nu_{ij}$ the complex phase associated to the next-nearest hopping, with $\nu_{ij}=\pm1$ as before. We set $\lambda_H=0.3 t$, as in the previous Section~\footnote{Both in the tight binding calculation and in the low energy effective theory, a change of $\lambda_H$ renormalizes the Fermi velocity of the edge modes in the bearded and zigzag cases, hence slightly modifying the size of the energy gaps. In the zigzag-armchair case, the location of the gap is also slightly shifted.}. The results of the different diagonalizations are shown in Fig.~\ref{fig:bands}: In each plot one can see the bulk bands above and below the bulk gap, which is crossed by two bands corresponding to counter propagating modes localized on the two edges of the strip.

Two facts need to be underlined: First, the Brillouin zone of the armchair strip is wide $\frac{2\pi}{\sqrt{3}a}$, while the ones obtained for the strips with bearded and zigzag edges have a width of $\frac{2\pi}{a}$. This is due to the fact that armchair edges have different periodicity with respect to the other two: more precisely the armchair edge has a period of $\sqrt{3}a$, while the others have a period of $a$, so that they are incommensurable. This fact is in close connection to the quasicrystalline structure studied in the preceeding Section. Secondarily, while both the bands obtained for the armchair and bearded strip have the Dirac point for $k=0$, the bands of the zigzag strip have it at $k=\frac{\pi}{a}$. The facts just highlighted have important consequences on the edge spectrum of the coupled bilayer system, that we now proceed to discuss.

\subsection{Approximate bands of the bilayer}
We would now like to discuss the properties of the edge spectrum of the bilayer system. In other words, we want to assess the previous square bilayer far form the vertices. Qualitatively, this is equivalent to studying the spectral properties of a strip of the quasicrystalline bilayer. As a starting point we analyze the uncoupled case ($\lambda_\perp=0$); We will then introduce the interlayer coupling in a perturbative fashion through an effective theory.

For a given combination of the edges of the upper and the lower layers, the bands of each bilayer strip should be obtainable by superimposing the bands of the two individual strips in an appropriate Brillouin zone, determined by the approximate periodicity of the bilayer strip. Indeed, since the period of the strip with armchair edges is incommensurable with the one of the strips with bearded or zigzag edges, the reciprocal space is not even defined for the bilayer strip. So, one has to take a sufficiently large supercell to approximate the bilayer strip as the periodic repetition of the chosen supercell. A bigger supercell results in a better approximation. Actually, this corresponds to approximate $\sqrt{3}$ as a ratio of integer numbers, let's say $p/q$: then $pa\ (\approx q\sqrt{3}a)$ will give the length of the supercell.

In this framework, we find an approximate Brillouin zone with a width of $\frac{2\pi}{pa}\ \left(\approx \frac{2\pi}{q\sqrt{3}a}\right)$, were the bands of the zigzag (bearded) strip are folded $p$ times and those of the armchair strip are folded $q$ times. It can be shown that this description in the folded Brillouin zone is perfectly equivalent to a corresponding one in an extended zone scheme: in the latter, the bands of the bilayer strip are obtained from superimposing those of the zigzag (bearded) and armchair strip, repeated respectively $q$ and $p$ times over an extended zone of width $q\frac{2\pi}{a}\ \left(\approx p\frac{2\pi}{\sqrt{3}a}\right)$. In the following, we stick with the extended zone scheme, for reasons that will become clear in the next Subsection.

By using the \emph{quasiperiodic} approximation just described, it is possible to construct the approximate bands of the uncoupled bilayer edge, and from those to predict where the gaps should open once the interlayer coupling is introduced. 

\begin{figure}
    \centering
    \includegraphics[width=\columnwidth]{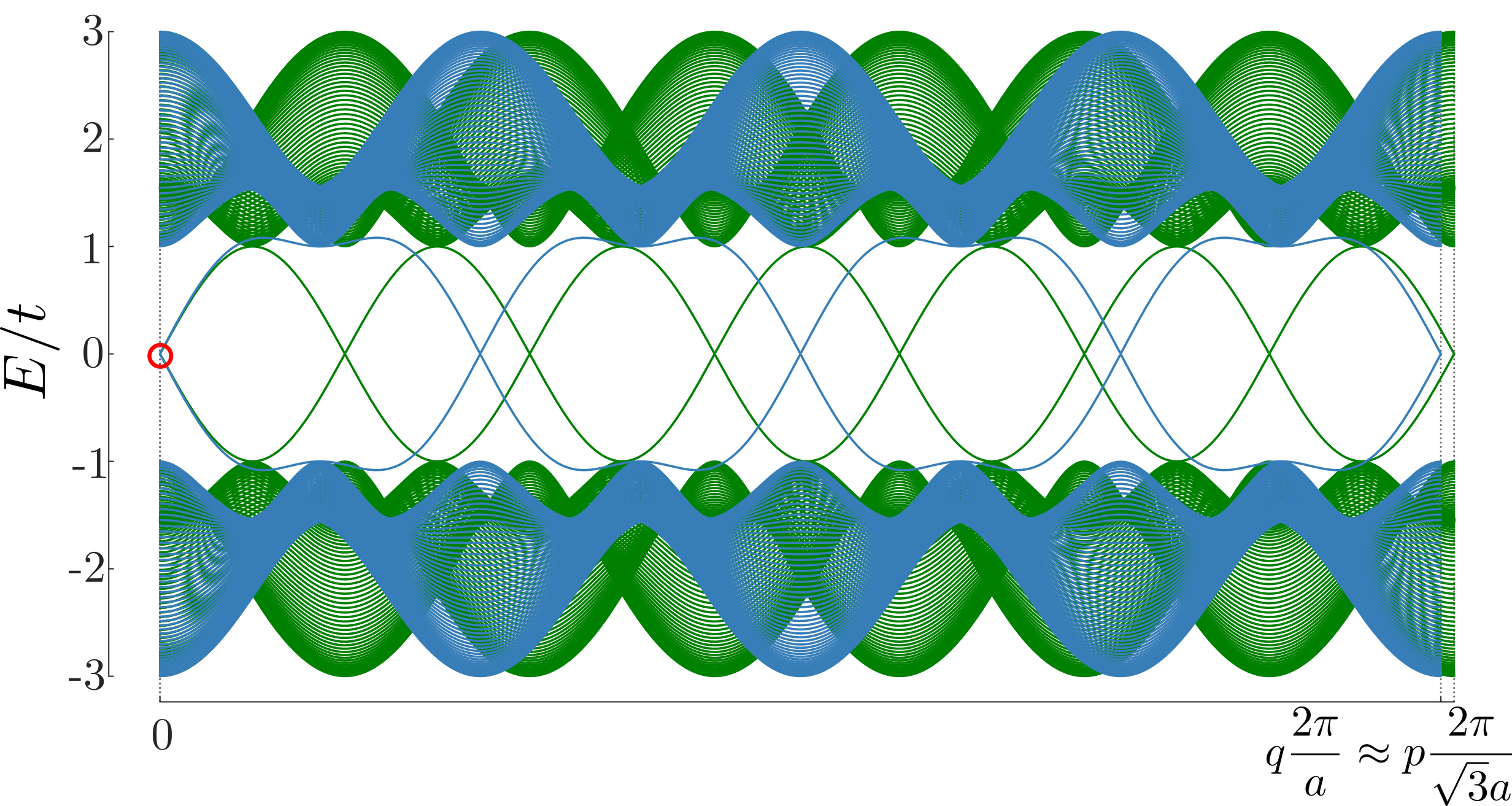}
    \caption{Bands of a bilayer strip for the combination of edges armchair-bearded, obtained setting $\lambda_\perp=0.3t$. The bands are drawn in an extended zone scheme, taking the supercell corresponding to the approximation $\sqrt{3}\approx 7/4$. The bands of the armchair strip are colored in green, while those of the bearded strip are colored in blue; the edge modes are clearly visible inside the bulk gap. The first crossing between the bands of the two edges occurs at $k=0$ and $E/t=0$ (red open circle).}
    \label{fig:bands_arm_dang}
\end{figure}

\begin{figure}
    \centering
    \includegraphics[width=\columnwidth]{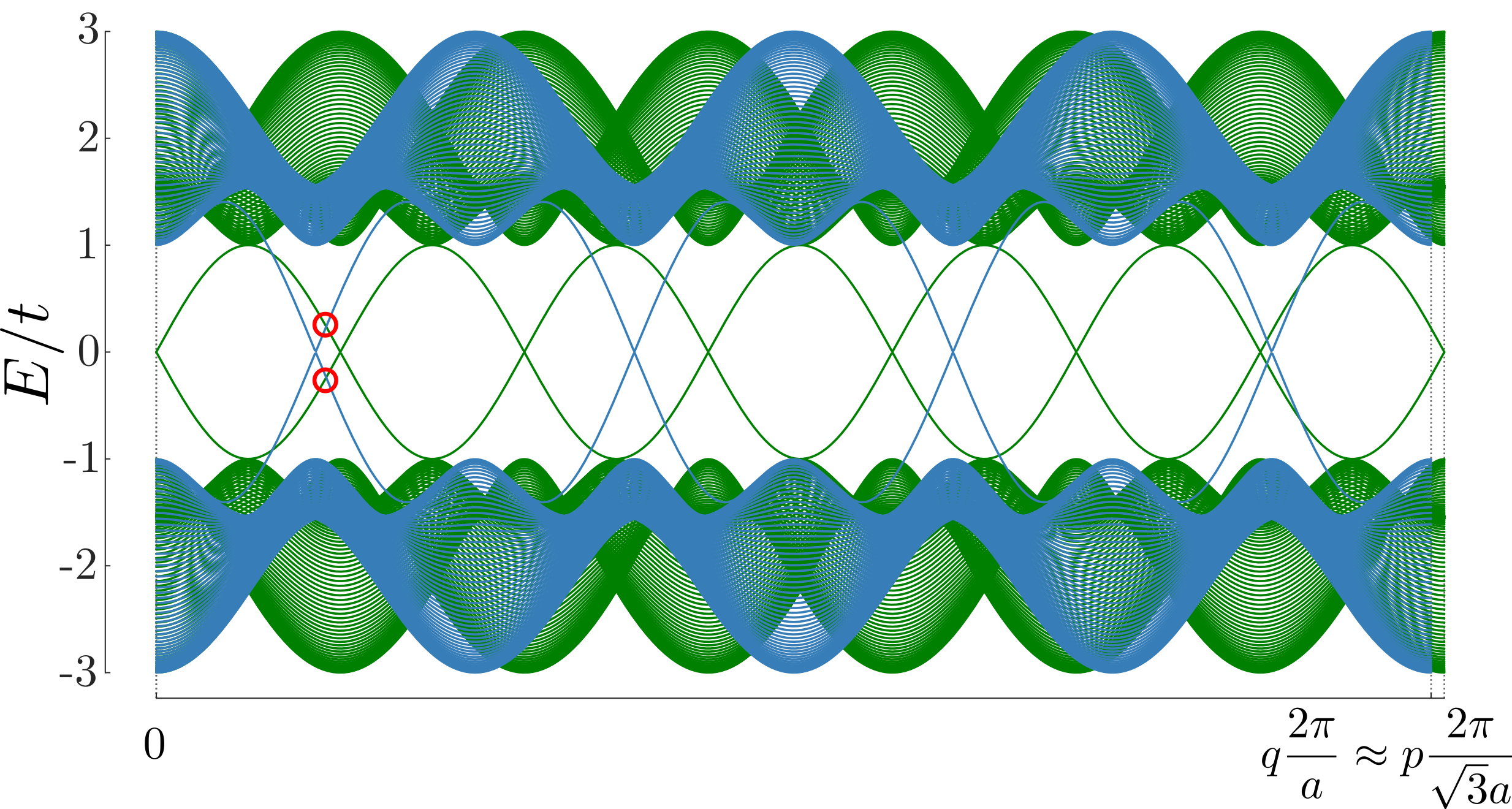}
    \caption{Bands of a bilayer strip for the combination of edges armchair-zigzag, obtained setting $\lambda_\perp=0.3t$. The bands are drawn in and extended zone scheme, taking the supercell corresponding to the approximation $\sqrt{3}\approx 7/4$. The bands of the armchair strip are colored in green, while those of the zigzag strip are colored in blue; the edge modes are clearly visible inside the bulk gap. The first crossing between the bands of the two edges occurs at $k\neq0$, around $E/t\approx \pm0.2$ (red open circles).}
    \label{fig:bands_arm_zig}
\end{figure}

In Fig.~\ref{fig:bands_arm_dang} (Fig.~\ref{fig:bands_arm_zig}) we show the superposition of the Haldane bands of an armchair strip, repeated over seven Brillouin zones, with those of a bearded (zigzag) strip, repeated over four Brillouin zones. As we discussed above, these are just the bands that one would obtain by considering $7/4$ as a rational approximant of $\sqrt{3}$ and imposing periodic boundary conditions on the corresponding supercell, in an extended zone scheme. In what follows, we will be interested only with the bands associated to the edge modes, which are those that run inside the bulk gap. The bulk bands of the strips are added for easier confrontation with Fig.~\ref{fig:bands}.

The introduction of the interlayer hopping causes gap opening between the edge bands of the two layers. More precisely, the gaps open in correspondence to the points of Fig.~\ref{fig:bands_arm_dang} and Fig.~\ref{fig:bands_arm_zig} where the bands of the two edges cross with each other, indicated with red circles in the figures. It is self evident that by considering higher order approximants of $\sqrt{3}$ (that come with greater values of $p$ and $q$), in the extended zone scheme the bands of the two single edges would be repeated many more times and this, due to incommensurability, would lead to more anti-crossing points, with each of them located at a different energy.

Anyway, a hierarchy exists between the different gaps, due to the fact that the interlayer hopping has not the same intensity for all $q$s in the reciprocal space. Indeed, we will now show that the intensity of the interlayer coupling reduces for increasing values of $q$ in the extended zone scheme of Fig.~\ref{fig:bands_arm_dang} and Fig.~\ref{fig:bands_arm_zig}. In light of this, we understand that the crucial observation of this Subsection is that in the armchair-bearded case there is the possibility to open a substantial gap at $q=0$ and hence zero energy, while this is not the case in the armchair-zigzag case. We will return on this later on.

\subsection{Continuum model for the interlayer hopping}
We now want to qualitatively analyze the amplitude of the gaps opening at the points of the spectra in Figs.~\ref{fig:bands_arm_dang} and \ref{fig:bands_arm_zig} where the bands of the two edges cross.
To do so, we change perspective a little bit. Guided by the fact that we are only interested in understanding how the introduction of the interlayer hopping affects the edge modes, we start by considering a model composed of two coupled 1-dimensional (1D) systems. In this new framework, we first develop a continuum model of the coupling term between the chains. Then we enforce the obtained result with some physical input from the problem at hand, that is the explicit form of the inter-chain coupling- that will be deduced from Eq.~(\ref{eq:t_perp}) -and the incommensurability between the two chains. As we will see, this approach will give us useful insight about the hierarchy among the different gaps that we anticipated above.

So, for the first part of the derivation we consider two generic 1D chains, with $N_1$ and $N_2$ cells respectively, coupled together by a generic Hamiltonian that will be denoted as $H_\perp$.  With the perturbation theory framework in mind, we want to calculate the matrix element~\cite{Bistritzer12233, PhysRevB.100.035101}
\begin{equation}
    T_{k_1 k_2}^{\alpha\beta}= \braket{\psi_{k_1\alpha}^{(1)}|H_{\perp}|\psi_{k_2\beta}^{(2)}},
    \label{eq:scattering_matrix}
\end{equation}
with the Bloch states $\psi_{k_1\alpha}^{(1)}$ and $\psi_{k_2\beta}^{(2)}$ given by
\begin{eqnarray}
    \ket{\psi_{k_1\alpha}^{(1)}} &= \dfrac{1}{\sqrt{N_1}} \sum_{n}e^{ik_1(n\Pi_1+\xi_\alpha)}\ket{n,\alpha},
    \label{eq:bloch_1}\\
    \ket{\psi_{k_2\beta}^{(2)}} &= \dfrac{1}{\sqrt{N_2}} \sum_{m}e^{ik_2(m\Pi_2+\xi_\beta)}\ket{m,\beta},
    \label{eq:bloch_2}
\end{eqnarray}
where $\Pi_{1,2}$ is the lattice period of the two chains, $\xi_{\alpha,\beta}$ is the positions of the atoms in the unit cell of chain $1,2$ and $k_1$ and $k_2$ are the respective Bloch momentum. Moreover $\ket{n,\alpha}$ ($\ket{m,\beta}$) is a Wannier function localized on the site at position $\xi_\alpha$ ($\xi_\beta$) in the $n$-th ($m$-th) cell of chain 1 (2). Inserting Eqs.~\ref{eq:bloch_1}  and~\ref{eq:bloch_2} into Eq.~\ref{eq:scattering_matrix} we have
\begin{multline}
    T_{k_1 k_2}^{\alpha\beta} =
    \dfrac{1}{\sqrt{N_1N_2}} \times\\ \sum_{n,m}e^{-ik_1(n\Pi_1+\xi_\alpha)+ik_2(m\Pi_2+\xi_\beta)}\braket{n,\alpha|H_{\perp}|m,\beta}.
\end{multline}
To proceed further, we make the so called \emph{two-center approximation}~\cite{Bistritzer12233}- indeed valid in the present case -that
\begin{equation}
    \braket{n,\alpha|H_{\perp}|m,\beta}\approx t(n\Pi_1+\xi_\alpha-m\Pi_2-\xi_\beta),
\end{equation}
where $t(x)$ is a function that will be specified later, and we use the Poisson summation formula
\begin{multline}
    \sum_{n,m}f(n\Pi_1+\xi_\alpha,m\Pi_2+\xi_\beta)\\
    =\dfrac{1}{\Pi_1\Pi_2}\sum_{l,j}\hat{f}\left(\frac{2\pi}{\Pi_1}l,\frac{2\pi}{\Pi_2}j\right) e^{i\frac{2\pi}{\Pi_1}l\xi_\alpha}e^{i\frac{2\pi}{\Pi_2}j\xi_\beta},
\end{multline}
adapted from~\cite{PhysRevB.100.035101} for the 1D case, where
\begin{equation}
    \hat{f}(q_1,q_2) = \int \mathrm{d}r_1\mathrm{d}r_2 e^{-iq_1r_1}e^{-iq_2r_2}f(r_1,r_2).
\end{equation}

After some algebra we find (modulo a multiplicative constant)
\begin{equation}
    T_{k_1 k_2}^{\alpha\beta} \sim \sum_{l,j}e^{i(\frac{2\pi}{\Pi_1}l\xi_\alpha-\frac{2\pi}{\Pi_2}j\xi_\beta)}\hat t\left(k_2+\frac{2\pi}{\Pi_2}j\right)\delta_{k_2+\frac{2\pi}{\Pi_2}j,k_1+\frac{2\pi}{\Pi_1}l},
    \label{eq:matrix_tunneling}
\end{equation}
where
\begin{equation}
    \hat{t}(k) = \int \mathrm{d} x e^{ikx} t(x).
\end{equation}
It is worth noting that Eq.~(\ref{eq:matrix_tunneling}) automatically enforces momentum conservation, modulo the reciprocal lattice vectors of the two chains.

Let's now introduce the physical input: We consider the sites on one edge of the bilayer GQ-like lattice and we call $x$ the coordinate along the edge.
Moreover, we assume the edge sites on the two layers to be perfectly lined up if looked at from a top/bottom view. Then from Eq.~(\ref{eq:t_perp}) applied only to sites on the chosen edge, we have that
\begin{equation}
    t(x) = t^\perp \exp \left(-\frac{\sqrt{x^2+d_0^2}-d_{0}}{\delta}\right).
    \label{eq:int_coupling_edge}
\end{equation}

\begin{figure}[hbt]
    \centering
    \includegraphics[width=\columnwidth]{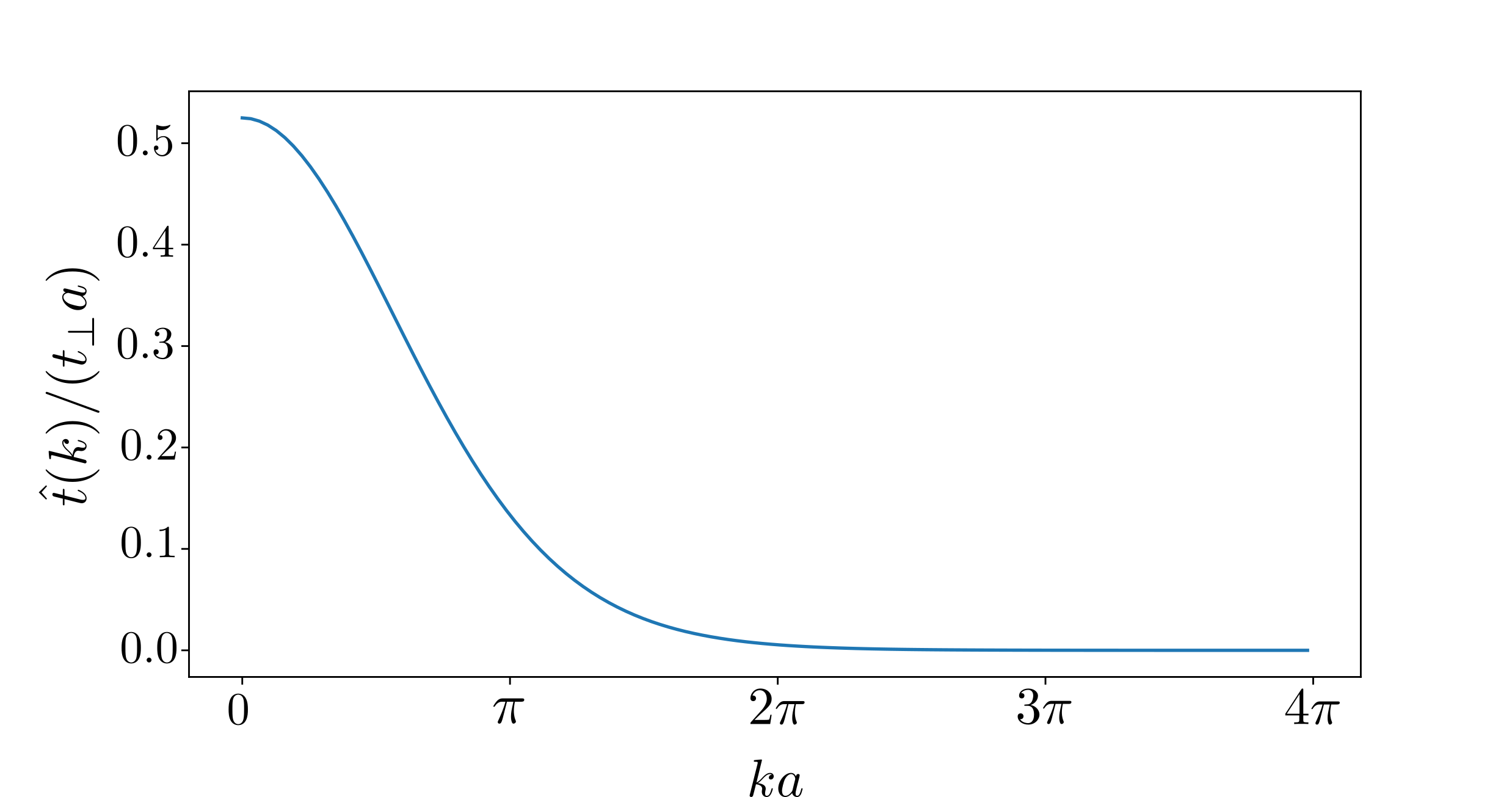}
    \caption{The Fourier transform along the edge of the interlayer coupling in Eq.~(\ref{eq:int_coupling_edge}), normalized to $t_\perp$. We set the parameters as described above in the text.}
    \label{fig:fourier_transform_tp}
\end{figure}

The Fourier transform of the preceding function cannot be computed analytically, but can be performed numerically once the parameters are set. The plot of $\hat{t}(k)/t_{\perp}$ is shown in Fig.~\ref{fig:fourier_transform_tp}: one can clearly see that for $|k|>\frac{2\pi}{a}$ the interlayer coupling becomes negligible.

Finally, we add the information that $\Pi_1$ and $\Pi_2$ are incommensurate: if this is the case, due to momentum conservation, once $k_1$ and $k_2$ are fixed in the respective reciprocal space, the sum in Eq.~(\ref{eq:matrix_tunneling}) reduces to one non-zero term at most. This leads us to a major simplification: indeed if we define $k\equiv k_2+\frac{2\pi}{\Pi_2}j=k_1+\frac{2\pi}{\Pi_1}l$, we can rewrite Eq.~(\ref{eq:matrix_tunneling}) as
\begin{equation}
    T_{k_1 k_2}^{\alpha\beta}\sim \hat{t}(k).
    \label{eq:final_matrix_element}
\end{equation}

Then, the decreasing trend of $\hat{t}(k)$ together with Eq.~(\ref{eq:final_matrix_element}) lead us to the following conclusion: the matrix elements in Eq.~(\ref{eq:scattering_matrix}), obtained for $k_1=k_2=k$, become less and less relevant as $|k|$ increases. As a consequence, with reference to Fig.~\ref{fig:bands_arm_dang} and Fig.~\ref{fig:bands_arm_zig}, the width of the gaps that open up in correspondence to the band crossings will decrease with increasing values of $|k|$. This statement makes us able to predict that the main gap in the case of armchair-bearded edges will open at $E/t\approx0$, while in the case of armchair-zigzag edges it will open at $E/t\approx\pm 0.2$. Indeed these are the positions in energy of the lowest $k$ band-crossings in Fig.~\ref{fig:bands_arm_dang} and Fig.~\ref{fig:bands_arm_zig} respectively, were they are indicated with red circles. We can also correctly deduce that the gaps that open in the armchair-zigzag case will be much smaller than the one that opens in the armchair-zigzag case. These predictions perfectly match the gap structure of the low energy spectra obtained for the finite size square samples reported in Fig.~\ref{fig:spect_dang} and Fig.~\ref{fig:spect_zig}.

\section{\label{sec:conc} Conclusion}
In this paper, we have analyzed the dependence on the edge structure of the topological modes characterizing a SOTI based on the GQ lattice. We have shown that, when the structure is composed of zigzag and armchair edges, the topological modes do not appear at the charge neutrality point. This fact, that might have consequences on the observability of the SOTI phase, has a very clear physical origin: The momentum mismatch between the topological modes of the two edges. This observation, together with the recognition that the Brillouin zones related to the armchair and the zigzag structures are incommensurate, have implications that go beyond the description of the SOTI phase, and implies the existence of one dimensional structures with fractal energy dispersion in systems where the topological states are hosted by armchair and zigzag edges in close proximity to each other.

\begin{acknowledgments}
We thank Stephen Spurrier, Lorenzo Privitera and Alberto Sciaccaluga for useful discussions. This work was supported by the ``Dipartimento di Eccellenza MIUR 2018-2022''.
\end{acknowledgments}

\appendix
\section{The dodecagonal case}
In Sec.~\ref{subsec:lattice}, we argued that the results presented would have remained qualitatively the same if one were to consider a dodecagonal sample instead of a square one. Indeeed, the low energy theory we developed in Sec.~\ref{sec:interpretation} refers to a single edge of the bilayer system, so that it can be applied to the case of dodecagonal shape as well. More explicitly, we expect that if we choose a dodecagonal sample with zigzag-armchair edges we are going to find two gaps opening around $E/t\approx\pm 0.2$, while if we choose a dodecagonal sample with bearded-armchair edges- as done in \cite{PhysRevResearch.2.033071} -we will find a single bigger gap opening around $E/t\approx 0$. This is indeed what happens.

\begin{figure}
    \centering
    \includegraphics[width=\linewidth]{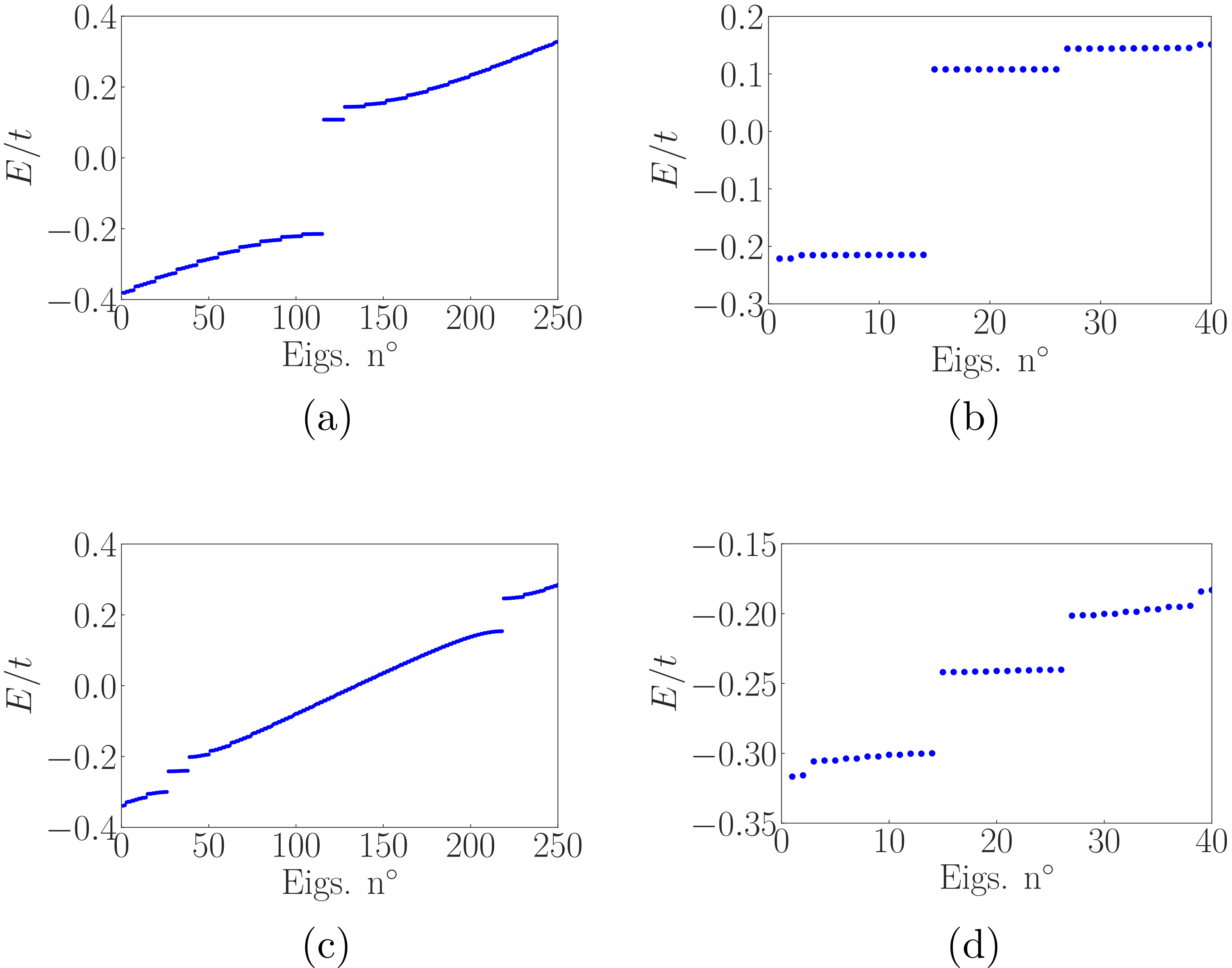}
    \caption{Low energy eigenvalues obtained by diagonalizing the Hamiltonian in Eq.~(\ref{eq:spurrier_ham}) for a dodecagonal sample with bearded-armchair edges (Panel (a)) and with zigzag-armchair edges (Panel (c)), setting $\lambda_\perp=2$ and $\lambda_H=0.3t$. The samples considered have apothems of length $278 a_{cc}$ and $326.5 a_{cc}$ respectively. In Panels (b) and (c) a zoom of the twelve degenerate in-gap eigenvalues of Panel (a) and (c) respectively.}
    \label{fig:appendix}
\end{figure}

In Fig.~\ref{fig:appendix} are reported the low energy eigenvalues obtained by diagonalizing the Hamiltonian in Eq.~(\ref{eq:spurrier_ham}) for finite-size dodecagonal sample, both with bearded-armchair edges and with zigzag-armchair edges. One can clearly see that in the bearded-armchair case a single gap opens in the edge spectrum, located close to the charge neutrality point (Fig.\ref{fig:appendix}(a)). This gap hosts twelve degenerate eigenvalues, as can be seen from the zoom in Fig.~\ref{fig:appendix}(b), corresponding to as many corner modes (probability densities not shown). On the other hand in the zigzag-armchair case, just as already shown for square samples, two smaller gaps open (Fig.\ref{fig:appendix}(c)). These are located away from charge neutrality point, at $E/t\approx\pm 0.2$, and the one at lower energy hosts twelve degenerate eigenvalues (zoom in Fig.\ref{fig:appendix}(d)), again corresponding to corner modes.

It is worth pointing out, that being the ratio between the edge length and the total number of sites signficantly smaller for dodecagonal samples than it is for square ones, in order to avoid hybridization between consecutive corners much bigger samples must be considered. This is not advantageous from a computational point of view and it is the reason we decided to deal with square samples in the paper.

\providecommand{\noopsort}[1]{}\providecommand{\singleletter}[1]{#1}%


\begin{thebibliography}{41}%
\makeatletter
\providecommand \@ifxundefined [1]{%
 \@ifx{#1\undefined}
}%
\providecommand \@ifnum [1]{%
 \ifnum #1\expandafter \@firstoftwo
 \else \expandafter \@secondoftwo
 \fi
}%
\providecommand \@ifx [1]{%
 \ifx #1\expandafter \@firstoftwo
 \else \expandafter \@secondoftwo
 \fi
}%
\providecommand \natexlab [1]{#1}%
\providecommand \enquote  [1]{``#1''}%
\providecommand \bibnamefont  [1]{#1}%
\providecommand \bibfnamefont [1]{#1}%
\providecommand \citenamefont [1]{#1}%
\providecommand \href@noop [0]{\@secondoftwo}%
\providecommand \href [0]{\begingroup \@sanitize@url \@href}%
\providecommand \@href[1]{\@@startlink{#1}\@@href}%
\providecommand \@@href[1]{\endgroup#1\@@endlink}%
\providecommand \@sanitize@url [0]{\catcode `\\12\catcode `\$12\catcode
  `\&12\catcode `\#12\catcode `\^12\catcode `\_12\catcode `\%12\relax}%
\providecommand \@@startlink[1]{}%
\providecommand \@@endlink[0]{}%
\providecommand \url  [0]{\begingroup\@sanitize@url \@url }%
\providecommand \@url [1]{\endgroup\@href {#1}{\urlprefix }}%
\providecommand \urlprefix  [0]{URL }%
\providecommand \Eprint [0]{\href }%
\providecommand \doibase [0]{https://doi.org/}%
\providecommand \selectlanguage [0]{\@gobble}%
\providecommand \bibinfo  [0]{\@secondoftwo}%
\providecommand \bibfield  [0]{\@secondoftwo}%
\providecommand \translation [1]{[#1]}%
\providecommand \BibitemOpen [0]{}%
\providecommand \bibitemStop [0]{}%
\providecommand \bibitemNoStop [0]{.\EOS\space}%
\providecommand \EOS [0]{\spacefactor3000\relax}%
\providecommand \BibitemShut  [1]{\csname bibitem#1\endcsname}%
\let\auto@bib@innerbib\@empty
\bibitem [{\citenamefont {Hasan}\ and\ \citenamefont
  {Kane}(2010)}]{RevModPhys.82.3045}%
  \BibitemOpen
  \bibfield  {author} {\bibinfo {author} {\bibfnamefont {M.~Z.}\ \bibnamefont
  {Hasan}}\ and\ \bibinfo {author} {\bibfnamefont {C.~L.}\ \bibnamefont
  {Kane}},\ }\bibfield  {title} {\bibinfo {title} {Colloquium: Topological
  insulators},\ }\href {https://doi.org/10.1103/RevModPhys.82.3045} {\bibfield
  {journal} {\bibinfo  {journal} {Rev. Mod. Phys.}\ }\textbf {\bibinfo {volume}
  {82}},\ \bibinfo {pages} {3045} (\bibinfo {year} {2010})}\BibitemShut
  {NoStop}%
\bibitem [{\citenamefont {Fu}\ \emph {et~al.}(2007)\citenamefont {Fu},
  \citenamefont {Kane},\ and\ \citenamefont {Mele}}]{PhysRevLett.98.106803}%
  \BibitemOpen
  \bibfield  {author} {\bibinfo {author} {\bibfnamefont {L.}~\bibnamefont
  {Fu}}, \bibinfo {author} {\bibfnamefont {C.~L.}\ \bibnamefont {Kane}},\ and\
  \bibinfo {author} {\bibfnamefont {E.~J.}\ \bibnamefont {Mele}},\ }\bibfield
  {title} {\bibinfo {title} {Topological insulators in three dimensions},\
  }\href {https://doi.org/10.1103/PhysRevLett.98.106803} {\bibfield  {journal}
  {\bibinfo  {journal} {Phys. Rev. Lett.}\ }\textbf {\bibinfo {volume} {98}},\
  \bibinfo {pages} {106803} (\bibinfo {year} {2007})}\BibitemShut {NoStop}%
\bibitem [{\citenamefont {Bernevig}\ \emph {et~al.}(2006)\citenamefont
  {Bernevig}, \citenamefont {Hughes},\ and\ \citenamefont
  {Zhang}}]{doi:10.1126/science.1133734}%
  \BibitemOpen
  \bibfield  {author} {\bibinfo {author} {\bibfnamefont {B.~A.}\ \bibnamefont
  {Bernevig}}, \bibinfo {author} {\bibfnamefont {T.~L.}\ \bibnamefont
  {Hughes}},\ and\ \bibinfo {author} {\bibfnamefont {S.-C.}\ \bibnamefont
  {Zhang}},\ }\bibfield  {title} {\bibinfo {title} {Quantum spin hall effect
  and topological phase transition in hgte quantum wells},\ }\href
  {https://doi.org/10.1126/science.1133734} {\bibfield  {journal} {\bibinfo
  {journal} {Science}\ }\textbf {\bibinfo {volume} {314}},\ \bibinfo {pages}
  {1757} (\bibinfo {year} {2006})}\BibitemShut {NoStop}%
\bibitem [{\citenamefont {Zhang}\ \emph {et~al.}(2009)\citenamefont {Zhang},
  \citenamefont {Liu}, \citenamefont {Qi}, \citenamefont {Dai}, \citenamefont
  {Fang},\ and\ \citenamefont {Zhang}}]{Zhang:2009zzf}%
  \BibitemOpen
  \bibfield  {author} {\bibinfo {author} {\bibfnamefont {H.}~\bibnamefont
  {Zhang}}, \bibinfo {author} {\bibfnamefont {C.-X.}\ \bibnamefont {Liu}},
  \bibinfo {author} {\bibfnamefont {X.-L.}\ \bibnamefont {Qi}}, \bibinfo
  {author} {\bibfnamefont {X.}~\bibnamefont {Dai}}, \bibinfo {author}
  {\bibfnamefont {Z.}~\bibnamefont {Fang}},\ and\ \bibinfo {author}
  {\bibfnamefont {S.-C.}\ \bibnamefont {Zhang}},\ }\bibfield  {title} {\bibinfo
  {title} {{Topological insulators in Bi2Se3, Bi2Te3 and Sb2Te3 with a single
  Dirac cone on the surface}},\ }\href {https://doi.org/10.1038/nphys1270}
  {\bibfield  {journal} {\bibinfo  {journal} {Nature Phys.}\ }\textbf {\bibinfo
  {volume} {5}},\ \bibinfo {pages} {438} (\bibinfo {year} {2009})}\BibitemShut
  {NoStop}%
\bibitem [{\citenamefont {Bernevig}\ and\ \citenamefont
  {Hughes}(2013)}]{bernevig2013topological}%
  \BibitemOpen
  \bibfield  {author} {\bibinfo {author} {\bibfnamefont {B.}~\bibnamefont
  {Bernevig}}\ and\ \bibinfo {author} {\bibfnamefont {T.}~\bibnamefont
  {Hughes}},\ }\href {https://books.google.it/books?id=wOn7JHSSxrsC} {\emph
  {\bibinfo {title} {Topological Insulators and Topological Superconductors}}}\
  (\bibinfo  {publisher} {Princeton University Press},\ \bibinfo {year}
  {2013})\BibitemShut {NoStop}%
\bibitem [{\citenamefont {Kane}\ and\ \citenamefont
  {Mele}(2005{\natexlab{a}})}]{PhysRevLett.95.146802}%
  \BibitemOpen
  \bibfield  {author} {\bibinfo {author} {\bibfnamefont {C.~L.}\ \bibnamefont
  {Kane}}\ and\ \bibinfo {author} {\bibfnamefont {E.~J.}\ \bibnamefont
  {Mele}},\ }\bibfield  {title} {\bibinfo {title} {${Z}_{2}$ topological order
  and the quantum spin hall effect},\ }\href
  {https://doi.org/10.1103/PhysRevLett.95.146802} {\bibfield  {journal}
  {\bibinfo  {journal} {Phys. Rev. Lett.}\ }\textbf {\bibinfo {volume} {95}},\
  \bibinfo {pages} {146802} (\bibinfo {year} {2005}{\natexlab{a}})}\BibitemShut
  {NoStop}%
\bibitem [{\citenamefont {Kane}\ and\ \citenamefont
  {Mele}(2005{\natexlab{b}})}]{PhysRevLett.95.226801}%
  \BibitemOpen
  \bibfield  {author} {\bibinfo {author} {\bibfnamefont {C.~L.}\ \bibnamefont
  {Kane}}\ and\ \bibinfo {author} {\bibfnamefont {E.~J.}\ \bibnamefont
  {Mele}},\ }\bibfield  {title} {\bibinfo {title} {Quantum spin hall effect in
  graphene},\ }\href {https://doi.org/10.1103/PhysRevLett.95.226801} {\bibfield
   {journal} {\bibinfo  {journal} {Phys. Rev. Lett.}\ }\textbf {\bibinfo
  {volume} {95}},\ \bibinfo {pages} {226801} (\bibinfo {year}
  {2005}{\natexlab{b}})}\BibitemShut {NoStop}%
\bibitem [{\citenamefont {He}\ \emph {et~al.}(2019)\citenamefont {He},
  \citenamefont {Sun},\ and\ \citenamefont {He}}]{He2019}%
  \BibitemOpen
  \bibfield  {author} {\bibinfo {author} {\bibfnamefont {M.}~\bibnamefont
  {He}}, \bibinfo {author} {\bibfnamefont {H.}~\bibnamefont {Sun}},\ and\
  \bibinfo {author} {\bibfnamefont {Q.~L.}\ \bibnamefont {He}},\ }\bibfield
  {title} {\bibinfo {title} {Topological insulator: Spintronics and quantum
  computations},\ }\href {https://doi.org/10.1007/s11467-019-0893-4} {\bibfield
   {journal} {\bibinfo  {journal} {Frontiers of Physics}\ }\textbf {\bibinfo
  {volume} {14}},\ \bibinfo {pages} {43401} (\bibinfo {year}
  {2019})}\BibitemShut {NoStop}%
\bibitem [{\citenamefont {Hsieh}\ \emph {et~al.}(2009)\citenamefont {Hsieh},
  \citenamefont {Xia}, \citenamefont {Qian}, \citenamefont {Wray},
  \citenamefont {Dil}, \citenamefont {Meier}, \citenamefont {Osterwalder},
  \citenamefont {Patthey}, \citenamefont {Checkelsky}, \citenamefont {Ong},
  \citenamefont {Fedorov}, \citenamefont {Lin}, \citenamefont {Bansil},
  \citenamefont {Grauer}, \citenamefont {Hor}, \citenamefont {Cava},\ and\
  \citenamefont {Hasan}}]{Hsieh2009}%
  \BibitemOpen
  \bibfield  {author} {\bibinfo {author} {\bibfnamefont {D.}~\bibnamefont
  {Hsieh}}, \bibinfo {author} {\bibfnamefont {Y.}~\bibnamefont {Xia}}, \bibinfo
  {author} {\bibfnamefont {D.}~\bibnamefont {Qian}}, \bibinfo {author}
  {\bibfnamefont {L.}~\bibnamefont {Wray}}, \bibinfo {author} {\bibfnamefont
  {J.~H.}\ \bibnamefont {Dil}}, \bibinfo {author} {\bibfnamefont
  {F.}~\bibnamefont {Meier}}, \bibinfo {author} {\bibfnamefont
  {J.}~\bibnamefont {Osterwalder}}, \bibinfo {author} {\bibfnamefont
  {L.}~\bibnamefont {Patthey}}, \bibinfo {author} {\bibfnamefont {J.~G.}\
  \bibnamefont {Checkelsky}}, \bibinfo {author} {\bibfnamefont {N.~P.}\
  \bibnamefont {Ong}}, \bibinfo {author} {\bibfnamefont {A.~V.}\ \bibnamefont
  {Fedorov}}, \bibinfo {author} {\bibfnamefont {H.}~\bibnamefont {Lin}},
  \bibinfo {author} {\bibfnamefont {A.}~\bibnamefont {Bansil}}, \bibinfo
  {author} {\bibfnamefont {D.}~\bibnamefont {Grauer}}, \bibinfo {author}
  {\bibfnamefont {Y.~S.}\ \bibnamefont {Hor}}, \bibinfo {author} {\bibfnamefont
  {R.~J.}\ \bibnamefont {Cava}},\ and\ \bibinfo {author} {\bibfnamefont
  {M.~Z.}\ \bibnamefont {Hasan}},\ }\bibfield  {title} {\bibinfo {title} {A
  tunable topological insulator in the spin helical dirac transport regime},\
  }\href {https://doi.org/10.1038/nature08234} {\bibfield  {journal} {\bibinfo
  {journal} {Nature}\ }\textbf {\bibinfo {volume} {460}},\ \bibinfo {pages}
  {1101} (\bibinfo {year} {2009})}\BibitemShut {NoStop}%
\bibitem [{\citenamefont {Mellnik}\ \emph {et~al.}(2014)\citenamefont
  {Mellnik}, \citenamefont {Lee}, \citenamefont {Richardella}, \citenamefont
  {Grab}, \citenamefont {Mintun}, \citenamefont {Fischer}, \citenamefont
  {Vaezi}, \citenamefont {Manchon}, \citenamefont {Kim}, \citenamefont
  {Samarth},\ and\ \citenamefont {Ralph}}]{Mellnik2014}%
  \BibitemOpen
  \bibfield  {author} {\bibinfo {author} {\bibfnamefont {A.~R.}\ \bibnamefont
  {Mellnik}}, \bibinfo {author} {\bibfnamefont {J.~S.}\ \bibnamefont {Lee}},
  \bibinfo {author} {\bibfnamefont {A.}~\bibnamefont {Richardella}}, \bibinfo
  {author} {\bibfnamefont {J.~L.}\ \bibnamefont {Grab}}, \bibinfo {author}
  {\bibfnamefont {P.~J.}\ \bibnamefont {Mintun}}, \bibinfo {author}
  {\bibfnamefont {M.~H.}\ \bibnamefont {Fischer}}, \bibinfo {author}
  {\bibfnamefont {A.}~\bibnamefont {Vaezi}}, \bibinfo {author} {\bibfnamefont
  {A.}~\bibnamefont {Manchon}}, \bibinfo {author} {\bibfnamefont {E.-A.}\
  \bibnamefont {Kim}}, \bibinfo {author} {\bibfnamefont {N.}~\bibnamefont
  {Samarth}},\ and\ \bibinfo {author} {\bibfnamefont {D.~C.}\ \bibnamefont
  {Ralph}},\ }\bibfield  {title} {\bibinfo {title} {Spin-transfer torque
  generated by a topological insulator},\ }\href
  {https://doi.org/10.1038/nature13534} {\bibfield  {journal} {\bibinfo
  {journal} {Nature}\ }\textbf {\bibinfo {volume} {511}},\ \bibinfo {pages}
  {449} (\bibinfo {year} {2014})}\BibitemShut {NoStop}%
\bibitem [{\citenamefont {Schindler}\ \emph
  {et~al.}(2018{\natexlab{a}})\citenamefont {Schindler}, \citenamefont {Cook},
  \citenamefont {Vergniory}, \citenamefont {Wang}, \citenamefont {Parkin},
  \citenamefont {Bernevig},\ and\ \citenamefont
  {Neupert}}]{doi:10.1126/sciadv.aat0346}%
  \BibitemOpen
  \bibfield  {author} {\bibinfo {author} {\bibfnamefont {F.}~\bibnamefont
  {Schindler}}, \bibinfo {author} {\bibfnamefont {A.~M.}\ \bibnamefont {Cook}},
  \bibinfo {author} {\bibfnamefont {M.~G.}\ \bibnamefont {Vergniory}}, \bibinfo
  {author} {\bibfnamefont {Z.}~\bibnamefont {Wang}}, \bibinfo {author}
  {\bibfnamefont {S.~S.~P.}\ \bibnamefont {Parkin}}, \bibinfo {author}
  {\bibfnamefont {B.~A.}\ \bibnamefont {Bernevig}},\ and\ \bibinfo {author}
  {\bibfnamefont {T.}~\bibnamefont {Neupert}},\ }\bibfield  {title} {\bibinfo
  {title} {Higher-order topological insulators},\ }\href
  {https://doi.org/10.1126/sciadv.aat0346} {\bibfield  {journal} {\bibinfo
  {journal} {Science Advances}\ }\textbf {\bibinfo {volume} {4}},\ \bibinfo
  {pages} {eaat0346} (\bibinfo {year} {2018}{\natexlab{a}})}\BibitemShut
  {NoStop}%
\bibitem [{\citenamefont {Benalcazar}\ \emph
  {et~al.}(2017{\natexlab{a}})\citenamefont {Benalcazar}, \citenamefont
  {Bernevig},\ and\ \citenamefont {Hughes}}]{doi:10.1126/science.aah6442}%
  \BibitemOpen
  \bibfield  {author} {\bibinfo {author} {\bibfnamefont {W.~A.}\ \bibnamefont
  {Benalcazar}}, \bibinfo {author} {\bibfnamefont {B.~A.}\ \bibnamefont
  {Bernevig}},\ and\ \bibinfo {author} {\bibfnamefont {T.~L.}\ \bibnamefont
  {Hughes}},\ }\bibfield  {title} {\bibinfo {title} {Quantized electric
  multipole insulators},\ }\href {https://doi.org/10.1126/science.aah6442}
  {\bibfield  {journal} {\bibinfo  {journal} {Science}\ }\textbf {\bibinfo
  {volume} {357}},\ \bibinfo {pages} {61} (\bibinfo {year}
  {2017}{\natexlab{a}})}\BibitemShut {NoStop}%
\bibitem [{\citenamefont {Benalcazar}\ \emph
  {et~al.}(2017{\natexlab{b}})\citenamefont {Benalcazar}, \citenamefont
  {Bernevig},\ and\ \citenamefont {Hughes}}]{PhysRevB.96.245115}%
  \BibitemOpen
  \bibfield  {author} {\bibinfo {author} {\bibfnamefont {W.~A.}\ \bibnamefont
  {Benalcazar}}, \bibinfo {author} {\bibfnamefont {B.~A.}\ \bibnamefont
  {Bernevig}},\ and\ \bibinfo {author} {\bibfnamefont {T.~L.}\ \bibnamefont
  {Hughes}},\ }\bibfield  {title} {\bibinfo {title} {Electric multipole
  moments, topological multipole moment pumping, and chiral hinge states in
  crystalline insulators},\ }\href {https://doi.org/10.1103/PhysRevB.96.245115}
  {\bibfield  {journal} {\bibinfo  {journal} {Phys. Rev. B}\ }\textbf {\bibinfo
  {volume} {96}},\ \bibinfo {pages} {245115} (\bibinfo {year}
  {2017}{\natexlab{b}})}\BibitemShut {NoStop}%
\bibitem [{\citenamefont {Langbehn}\ \emph {et~al.}(2017)\citenamefont
  {Langbehn}, \citenamefont {Peng}, \citenamefont {Trifunovic}, \citenamefont
  {von Oppen},\ and\ \citenamefont {Brouwer}}]{PhysRevLett.119.246401}%
  \BibitemOpen
  \bibfield  {author} {\bibinfo {author} {\bibfnamefont {J.}~\bibnamefont
  {Langbehn}}, \bibinfo {author} {\bibfnamefont {Y.}~\bibnamefont {Peng}},
  \bibinfo {author} {\bibfnamefont {L.}~\bibnamefont {Trifunovic}}, \bibinfo
  {author} {\bibfnamefont {F.}~\bibnamefont {von Oppen}},\ and\ \bibinfo
  {author} {\bibfnamefont {P.~W.}\ \bibnamefont {Brouwer}},\ }\bibfield
  {title} {\bibinfo {title} {Reflection-symmetric second-order topological
  insulators and superconductors},\ }\href
  {https://doi.org/10.1103/PhysRevLett.119.246401} {\bibfield  {journal}
  {\bibinfo  {journal} {Phys. Rev. Lett.}\ }\textbf {\bibinfo {volume} {119}},\
  \bibinfo {pages} {246401} (\bibinfo {year} {2017})}\BibitemShut {NoStop}%
\bibitem [{\citenamefont {Fang}\ and\ \citenamefont
  {Fu}(2019)}]{doi:10.1126/sciadv.aat2374}%
  \BibitemOpen
  \bibfield  {author} {\bibinfo {author} {\bibfnamefont {C.}~\bibnamefont
  {Fang}}\ and\ \bibinfo {author} {\bibfnamefont {L.}~\bibnamefont {Fu}},\
  }\bibfield  {title} {\bibinfo {title} {New classes of topological crystalline
  insulators having surface rotation anomaly},\ }\href
  {https://doi.org/10.1126/sciadv.aat2374} {\bibfield  {journal} {\bibinfo
  {journal} {Science Advances}\ }\textbf {\bibinfo {volume} {5}},\ \bibinfo
  {pages} {eaat2374} (\bibinfo {year} {2019})}\BibitemShut {NoStop}%
\bibitem [{\citenamefont {Noguchi}\ \emph {et~al.}(2021)\citenamefont
  {Noguchi}, \citenamefont {Kobayashi}, \citenamefont {Jiang}, \citenamefont
  {Kuroda}, \citenamefont {Takahashi}, \citenamefont {Xu}, \citenamefont {Lee},
  \citenamefont {Hirayama}, \citenamefont {Ochi}, \citenamefont {Shirasawa},
  \citenamefont {Zhang}, \citenamefont {Lin}, \citenamefont {Bareille},
  \citenamefont {Sakuragi}, \citenamefont {Tanaka}, \citenamefont {Kunisada},
  \citenamefont {Kurokawa}, \citenamefont {Yaji}, \citenamefont {Harasawa},\
  and\ \citenamefont {Kondo}}]{noguchiHoti}%
  \BibitemOpen
  \bibfield  {author} {\bibinfo {author} {\bibfnamefont {R.}~\bibnamefont
  {Noguchi}}, \bibinfo {author} {\bibfnamefont {M.}~\bibnamefont {Kobayashi}},
  \bibinfo {author} {\bibfnamefont {Z.}~\bibnamefont {Jiang}}, \bibinfo
  {author} {\bibfnamefont {K.}~\bibnamefont {Kuroda}}, \bibinfo {author}
  {\bibfnamefont {T.}~\bibnamefont {Takahashi}}, \bibinfo {author}
  {\bibfnamefont {Z.}~\bibnamefont {Xu}}, \bibinfo {author} {\bibfnamefont
  {D.}~\bibnamefont {Lee}}, \bibinfo {author} {\bibfnamefont {M.}~\bibnamefont
  {Hirayama}}, \bibinfo {author} {\bibfnamefont {M.}~\bibnamefont {Ochi}},
  \bibinfo {author} {\bibfnamefont {T.}~\bibnamefont {Shirasawa}}, \bibinfo
  {author} {\bibfnamefont {P.}~\bibnamefont {Zhang}}, \bibinfo {author}
  {\bibfnamefont {C.}~\bibnamefont {Lin}}, \bibinfo {author} {\bibfnamefont
  {C.}~\bibnamefont {Bareille}}, \bibinfo {author} {\bibfnamefont
  {S.}~\bibnamefont {Sakuragi}}, \bibinfo {author} {\bibfnamefont
  {H.}~\bibnamefont {Tanaka}}, \bibinfo {author} {\bibfnamefont
  {S.}~\bibnamefont {Kunisada}}, \bibinfo {author} {\bibfnamefont
  {K.}~\bibnamefont {Kurokawa}}, \bibinfo {author} {\bibfnamefont
  {K.}~\bibnamefont {Yaji}}, \bibinfo {author} {\bibfnamefont {A.}~\bibnamefont
  {Harasawa}},\ and\ \bibinfo {author} {\bibfnamefont {T.}~\bibnamefont
  {Kondo}},\ }\bibfield  {title} {\bibinfo {title} {Evidence for a higher-order
  topological insulator in a three-dimensional material built from van der
  waals stacking of bismuth-halide chains},\ }\href
  {https://doi.org/10.1038/s41563-020-00871-7} {\bibfield  {journal} {\bibinfo
  {journal} {Nature Materials}\ }\textbf {\bibinfo {volume} {20}},\ \bibinfo
  {pages} {1} (\bibinfo {year} {2021})}\BibitemShut {NoStop}%
\bibitem [{\citenamefont {Schindler}\ \emph
  {et~al.}(2018{\natexlab{b}})\citenamefont {Schindler}, \citenamefont {Wang},
  \citenamefont {Vergniory}, \citenamefont {Cook}, \citenamefont {Murani},
  \citenamefont {Sengupta}, \citenamefont {Kasumov}, \citenamefont {Deblock},
  \citenamefont {Jeon}, \citenamefont {Drozdov}, \citenamefont {Bouchiat},
  \citenamefont {Guéron}, \citenamefont {Yazdani}, \citenamefont {Bernevig},\
  and\ \citenamefont {Neupert}}]{schindlerHOTI}%
  \BibitemOpen
  \bibfield  {author} {\bibinfo {author} {\bibfnamefont {F.}~\bibnamefont
  {Schindler}}, \bibinfo {author} {\bibfnamefont {Z.}~\bibnamefont {Wang}},
  \bibinfo {author} {\bibfnamefont {M.}~\bibnamefont {Vergniory}}, \bibinfo
  {author} {\bibfnamefont {A.}~\bibnamefont {Cook}}, \bibinfo {author}
  {\bibfnamefont {A.}~\bibnamefont {Murani}}, \bibinfo {author} {\bibfnamefont
  {S.}~\bibnamefont {Sengupta}}, \bibinfo {author} {\bibfnamefont
  {A.}~\bibnamefont {Kasumov}}, \bibinfo {author} {\bibfnamefont
  {R.}~\bibnamefont {Deblock}}, \bibinfo {author} {\bibfnamefont
  {S.}~\bibnamefont {Jeon}}, \bibinfo {author} {\bibfnamefont {I.}~\bibnamefont
  {Drozdov}}, \bibinfo {author} {\bibfnamefont {H.}~\bibnamefont {Bouchiat}},
  \bibinfo {author} {\bibfnamefont {S.}~\bibnamefont {Guéron}}, \bibinfo
  {author} {\bibfnamefont {A.}~\bibnamefont {Yazdani}}, \bibinfo {author}
  {\bibfnamefont {B.}~\bibnamefont {Bernevig}},\ and\ \bibinfo {author}
  {\bibfnamefont {T.}~\bibnamefont {Neupert}},\ }\bibfield  {title} {\bibinfo
  {title} {Higher-order topology in bismuth},\ }\href
  {https://doi.org/10.1038/s41567-018-0224-7} {\bibfield  {journal} {\bibinfo
  {journal} {Nature Physics}\ }\textbf {\bibinfo {volume} {14}} (\bibinfo
  {year} {2018}{\natexlab{b}})}\BibitemShut {NoStop}%
\bibitem [{\citenamefont {Bindi}\ \emph {et~al.}(2009)\citenamefont {Bindi},
  \citenamefont {Steinhardt}, \citenamefont {Yao},\ and\ \citenamefont
  {Lu}}]{doi:10.1126/science.1170827}%
  \BibitemOpen
  \bibfield  {author} {\bibinfo {author} {\bibfnamefont {L.}~\bibnamefont
  {Bindi}}, \bibinfo {author} {\bibfnamefont {P.~J.}\ \bibnamefont
  {Steinhardt}}, \bibinfo {author} {\bibfnamefont {N.}~\bibnamefont {Yao}},\
  and\ \bibinfo {author} {\bibfnamefont {P.~J.}\ \bibnamefont {Lu}},\
  }\bibfield  {title} {\bibinfo {title} {Natural quasicrystals},\ }\href
  {https://doi.org/10.1126/science.1170827} {\bibfield  {journal} {\bibinfo
  {journal} {Science}\ }\textbf {\bibinfo {volume} {324}},\ \bibinfo {pages}
  {1306} (\bibinfo {year} {2009})}\BibitemShut {NoStop}%
\bibitem [{\citenamefont {Chen}\ \emph {et~al.}(2020)\citenamefont {Chen},
  \citenamefont {Chen}, \citenamefont {Gao}, \citenamefont {Zhou},\ and\
  \citenamefont {Xu}}]{PhysRevLett.124.036803}%
  \BibitemOpen
  \bibfield  {author} {\bibinfo {author} {\bibfnamefont {R.}~\bibnamefont
  {Chen}}, \bibinfo {author} {\bibfnamefont {C.-Z.}\ \bibnamefont {Chen}},
  \bibinfo {author} {\bibfnamefont {J.-H.}\ \bibnamefont {Gao}}, \bibinfo
  {author} {\bibfnamefont {B.}~\bibnamefont {Zhou}},\ and\ \bibinfo {author}
  {\bibfnamefont {D.-H.}\ \bibnamefont {Xu}},\ }\bibfield  {title} {\bibinfo
  {title} {Higher-order topological insulators in quasicrystals},\ }\href
  {https://doi.org/10.1103/PhysRevLett.124.036803} {\bibfield  {journal}
  {\bibinfo  {journal} {Phys. Rev. Lett.}\ }\textbf {\bibinfo {volume} {124}},\
  \bibinfo {pages} {036803} (\bibinfo {year} {2020})}\BibitemShut {NoStop}%
\bibitem [{\citenamefont {Varjas}\ \emph {et~al.}(2019)\citenamefont {Varjas},
  \citenamefont {Lau}, \citenamefont {P\"oyh\"onen}, \citenamefont {Akhmerov},
  \citenamefont {Pikulin},\ and\ \citenamefont
  {Fulga}}]{PhysRevLett.123.196401}%
  \BibitemOpen
  \bibfield  {author} {\bibinfo {author} {\bibfnamefont {D.}~\bibnamefont
  {Varjas}}, \bibinfo {author} {\bibfnamefont {A.}~\bibnamefont {Lau}},
  \bibinfo {author} {\bibfnamefont {K.}~\bibnamefont {P\"oyh\"onen}}, \bibinfo
  {author} {\bibfnamefont {A.~R.}\ \bibnamefont {Akhmerov}}, \bibinfo {author}
  {\bibfnamefont {D.~I.}\ \bibnamefont {Pikulin}},\ and\ \bibinfo {author}
  {\bibfnamefont {I.~C.}\ \bibnamefont {Fulga}},\ }\bibfield  {title} {\bibinfo
  {title} {Topological phases without crystalline counterparts},\ }\href
  {https://doi.org/10.1103/PhysRevLett.123.196401} {\bibfield  {journal}
  {\bibinfo  {journal} {Phys. Rev. Lett.}\ }\textbf {\bibinfo {volume} {123}},\
  \bibinfo {pages} {196401} (\bibinfo {year} {2019})}\BibitemShut {NoStop}%
\bibitem [{\citenamefont {Spurrier}\ and\ \citenamefont
  {Cooper}(2020)}]{PhysRevResearch.2.033071}%
  \BibitemOpen
  \bibfield  {author} {\bibinfo {author} {\bibfnamefont {S.}~\bibnamefont
  {Spurrier}}\ and\ \bibinfo {author} {\bibfnamefont {N.~R.}\ \bibnamefont
  {Cooper}},\ }\bibfield  {title} {\bibinfo {title} {Kane-mele with a twist:
  Quasicrystalline higher-order topological insulators with fractional mass
  kinks},\ }\href {https://doi.org/10.1103/PhysRevResearch.2.033071} {\bibfield
   {journal} {\bibinfo  {journal} {Phys. Rev. Research}\ }\textbf {\bibinfo
  {volume} {2}},\ \bibinfo {pages} {033071} (\bibinfo {year}
  {2020})}\BibitemShut {NoStop}%
\bibitem [{\citenamefont {McIver}\ \emph {et~al.}(2020)\citenamefont {McIver},
  \citenamefont {Schulte}, \citenamefont {Stein}, \citenamefont {Matsuyama},
  \citenamefont {Jotzu}, \citenamefont {Meier},\ and\ \citenamefont
  {Cavalleri}}]{maciver}%
  \BibitemOpen
  \bibfield  {author} {\bibinfo {author} {\bibfnamefont {J.~W.}\ \bibnamefont
  {McIver}}, \bibinfo {author} {\bibfnamefont {B.}~\bibnamefont {Schulte}},
  \bibinfo {author} {\bibfnamefont {F.-U.}\ \bibnamefont {Stein}}, \bibinfo
  {author} {\bibfnamefont {T.}~\bibnamefont {Matsuyama}}, \bibinfo {author}
  {\bibfnamefont {G.}~\bibnamefont {Jotzu}}, \bibinfo {author} {\bibfnamefont
  {G.}~\bibnamefont {Meier}},\ and\ \bibinfo {author} {\bibfnamefont
  {A.}~\bibnamefont {Cavalleri}},\ }\bibfield  {title} {\bibinfo {title}
  {Light-induced anomalous hall effect in graphene},\ }\href
  {https://doi.org/10.1038/s41567-019-0698-y} {\bibfield  {journal} {\bibinfo
  {journal} {Nature Physics}\ }\textbf {\bibinfo {volume} {16}},\ \bibinfo
  {pages} {38} (\bibinfo {year} {2020})}\BibitemShut {NoStop}%
\bibitem [{\citenamefont {Ahn}\ \emph {et~al.}(2018)\citenamefont {Ahn},
  \citenamefont {Moon}, \citenamefont {Kim}, \citenamefont {Kim}, \citenamefont
  {Shin}, \citenamefont {Kim}, \citenamefont {Cha}, \citenamefont {Kahng},
  \citenamefont {Kim}, \citenamefont {Koshino}, \citenamefont {Son},
  \citenamefont {Yang},\ and\ \citenamefont
  {Ahn}}]{doi:10.1126/science.aar8412}%
  \BibitemOpen
  \bibfield  {author} {\bibinfo {author} {\bibfnamefont {S.~J.}\ \bibnamefont
  {Ahn}}, \bibinfo {author} {\bibfnamefont {P.}~\bibnamefont {Moon}}, \bibinfo
  {author} {\bibfnamefont {T.-H.}\ \bibnamefont {Kim}}, \bibinfo {author}
  {\bibfnamefont {H.-W.}\ \bibnamefont {Kim}}, \bibinfo {author} {\bibfnamefont
  {H.-C.}\ \bibnamefont {Shin}}, \bibinfo {author} {\bibfnamefont {E.~H.}\
  \bibnamefont {Kim}}, \bibinfo {author} {\bibfnamefont {H.~W.}\ \bibnamefont
  {Cha}}, \bibinfo {author} {\bibfnamefont {S.-J.}\ \bibnamefont {Kahng}},
  \bibinfo {author} {\bibfnamefont {P.}~\bibnamefont {Kim}}, \bibinfo {author}
  {\bibfnamefont {M.}~\bibnamefont {Koshino}}, \bibinfo {author} {\bibfnamefont
  {Y.-W.}\ \bibnamefont {Son}}, \bibinfo {author} {\bibfnamefont {C.-W.}\
  \bibnamefont {Yang}},\ and\ \bibinfo {author} {\bibfnamefont {J.~R.}\
  \bibnamefont {Ahn}},\ }\bibfield  {title} {\bibinfo {title} {Dirac electrons
  in a dodecagonal graphene quasicrystal},\ }\href
  {https://doi.org/10.1126/science.aar8412} {\bibfield  {journal} {\bibinfo
  {journal} {Science}\ }\textbf {\bibinfo {volume} {361}},\ \bibinfo {pages}
  {782} (\bibinfo {year} {2018})}\BibitemShut {NoStop}%
\bibitem [{\citenamefont {St{\"u}hler}\ \emph {et~al.}(2022)\citenamefont
  {St{\"u}hler}, \citenamefont {Kowalewski}, \citenamefont {Reis},
  \citenamefont {Jungblut}, \citenamefont {Dominguez}, \citenamefont {Scharf},
  \citenamefont {Li}, \citenamefont {Sch{\"a}fer}, \citenamefont {Hankiewicz},\
  and\ \citenamefont {Claessen}}]{claessen}%
  \BibitemOpen
  \bibfield  {author} {\bibinfo {author} {\bibfnamefont {R.}~\bibnamefont
  {St{\"u}hler}}, \bibinfo {author} {\bibfnamefont {A.}~\bibnamefont
  {Kowalewski}}, \bibinfo {author} {\bibfnamefont {F.}~\bibnamefont {Reis}},
  \bibinfo {author} {\bibfnamefont {D.}~\bibnamefont {Jungblut}}, \bibinfo
  {author} {\bibfnamefont {F.}~\bibnamefont {Dominguez}}, \bibinfo {author}
  {\bibfnamefont {B.}~\bibnamefont {Scharf}}, \bibinfo {author} {\bibfnamefont
  {G.}~\bibnamefont {Li}}, \bibinfo {author} {\bibfnamefont {J.}~\bibnamefont
  {Sch{\"a}fer}}, \bibinfo {author} {\bibfnamefont {E.~M.}\ \bibnamefont
  {Hankiewicz}},\ and\ \bibinfo {author} {\bibfnamefont {R.}~\bibnamefont
  {Claessen}},\ }\bibfield  {title} {\bibinfo {title} {Effective lifting of the
  topological protection of quantum spin hall edge states by edge coupling},\
  }\href {https://doi.org/10.1038/s41467-022-30996-z} {\bibfield  {journal}
  {\bibinfo  {journal} {Nature Communications}\ }\textbf {\bibinfo {volume}
  {13}},\ \bibinfo {pages} {3480} (\bibinfo {year} {2022})}\BibitemShut
  {NoStop}%
\bibitem [{\citenamefont {Castro~Neto}\ \emph {et~al.}(2009)\citenamefont
  {Castro~Neto}, \citenamefont {Guinea}, \citenamefont {Peres}, \citenamefont
  {Novoselov},\ and\ \citenamefont {Geim}}]{RevModPhys.81.109}%
  \BibitemOpen
  \bibfield  {author} {\bibinfo {author} {\bibfnamefont {A.~H.}\ \bibnamefont
  {Castro~Neto}}, \bibinfo {author} {\bibfnamefont {F.}~\bibnamefont {Guinea}},
  \bibinfo {author} {\bibfnamefont {N.~M.~R.}\ \bibnamefont {Peres}}, \bibinfo
  {author} {\bibfnamefont {K.~S.}\ \bibnamefont {Novoselov}},\ and\ \bibinfo
  {author} {\bibfnamefont {A.~K.}\ \bibnamefont {Geim}},\ }\bibfield  {title}
  {\bibinfo {title} {The electronic properties of graphene},\ }\href
  {https://doi.org/10.1103/RevModPhys.81.109} {\bibfield  {journal} {\bibinfo
  {journal} {Rev. Mod. Phys.}\ }\textbf {\bibinfo {volume} {81}},\ \bibinfo
  {pages} {109} (\bibinfo {year} {2009})}\BibitemShut {NoStop}%
\bibitem [{\citenamefont {Cao}\ \emph {et~al.}(2017)\citenamefont {Cao},
  \citenamefont {Zhao},\ and\ \citenamefont {Louie}}]{PhysRevLett.119.076401}%
  \BibitemOpen
  \bibfield  {author} {\bibinfo {author} {\bibfnamefont {T.}~\bibnamefont
  {Cao}}, \bibinfo {author} {\bibfnamefont {F.}~\bibnamefont {Zhao}},\ and\
  \bibinfo {author} {\bibfnamefont {S.~G.}\ \bibnamefont {Louie}},\ }\bibfield
  {title} {\bibinfo {title} {Topological phases in graphene nanoribbons:
  Junction states, spin centers, and quantum spin chains},\ }\href
  {https://doi.org/10.1103/PhysRevLett.119.076401} {\bibfield  {journal}
  {\bibinfo  {journal} {Phys. Rev. Lett.}\ }\textbf {\bibinfo {volume} {119}},\
  \bibinfo {pages} {076401} (\bibinfo {year} {2017})}\BibitemShut {NoStop}%
\bibitem [{\citenamefont {Schaad}\ and\ \citenamefont
  {Stampfli}(2021)}]{schaad2021quasiperiodic}%
  \BibitemOpen
  \bibfield  {author} {\bibinfo {author} {\bibfnamefont {T.~P.}\ \bibnamefont
  {Schaad}}\ and\ \bibinfo {author} {\bibfnamefont {P.}~\bibnamefont
  {Stampfli}},\ }\href@noop {} {\bibinfo {title} {A quasiperiodic tiling with
  12-fold rotational symmetry and inflation factor $1 + \sqrt{3}$}} (\bibinfo
  {year} {2021}),\ \Eprint {https://arxiv.org/abs/2102.06046} {arXiv:2102.06046
  [math.HO]} \BibitemShut {NoStop}%
\bibitem [{\citenamefont {Altounian}\ and\ \citenamefont
  {Stampfli}(2018)}]{science_stampfli}%
  \BibitemOpen
  \bibfield  {author} {\bibinfo {author} {\bibfnamefont {V.}~\bibnamefont
  {Altounian}}\ and\ \bibinfo {author} {\bibfnamefont {P.}~\bibnamefont
  {Stampfli}},\ }\href
  {https://www.science.org/content/blog-post/cover-stories-making-graphene-quasicrystals-cover}
  {\bibinfo {title} {Cover stories: Making the graphene quasicrystals cover}}
  (\bibinfo {year} {2018})\BibitemShut {NoStop}%
\bibitem [{\citenamefont {Traverso}()}]{ncc}%
  \BibitemOpen
  \bibfield  {author} {\bibinfo {author} {\bibfnamefont {S.}~\bibnamefont
  {Traverso}},\ }\bibfield  {title} {\bibinfo {title} {Robustness of a
  quasicrystalline higher order topological insulator},\ }\href@noop {}
  {\bibinfo  {journal} {Il Nuovo Cimento C}\ \textbf
  {\bibinfo {volume} {45}},\ \bibinfo {pages} {1} (\bibinfo {year}
  {2022})}\BibitemShut
  {NoStop}%
\bibitem [{Note1()}]{Note1}%
  \BibitemOpen
\bibfield  {journal} {  }\bibinfo {note} {It is worth pointing out that this
  fact is true for a dodecagonal shape as well. This makes so that, as stated
  above, the following analysis can be replicated for samples with dodecagonal
  geometry.}\BibitemShut {Stop}%
\bibitem [{\citenamefont {Haldane}(1988)}]{PhysRevLett.61.2015}%
  \BibitemOpen
  \bibfield  {author} {\bibinfo {author} {\bibfnamefont {F.~D.~M.}\
  \bibnamefont {Haldane}},\ }\bibfield  {title} {\bibinfo {title} {Model for a
  quantum hall effect without landau levels: Condensed-matter realization of
  the ``parity anomaly''},\ }\href {https://doi.org/10.1103/PhysRevLett.61.2015}
  {\bibfield  {journal} {\bibinfo  {journal} {Phys. Rev. Lett.}\ }\textbf
  {\bibinfo {volume} {61}},\ \bibinfo {pages} {2015} (\bibinfo {year}
  {1988})}\BibitemShut {NoStop}%
\bibitem [{\citenamefont {Moon}\ and\ \citenamefont
  {Koshino}(2012)}]{PhysRevB.85.195458}%
  \BibitemOpen
  \bibfield  {author} {\bibinfo {author} {\bibfnamefont {P.}~\bibnamefont
  {Moon}}\ and\ \bibinfo {author} {\bibfnamefont {M.}~\bibnamefont {Koshino}},\
  }\bibfield  {title} {\bibinfo {title} {Energy spectrum and quantum hall
  effect in twisted bilayer graphene},\ }\href
  {https://doi.org/10.1103/PhysRevB.85.195458} {\bibfield  {journal} {\bibinfo
  {journal} {Phys. Rev. B}\ }\textbf {\bibinfo {volume} {85}},\ \bibinfo
  {pages} {195458} (\bibinfo {year} {2012})}\BibitemShut {NoStop}%
\bibitem [{\citenamefont {Moon}\ and\ \citenamefont
  {Koshino}(2013)}]{PhysRevB.87.205404}%
  \BibitemOpen
  \bibfield  {author} {\bibinfo {author} {\bibfnamefont {P.}~\bibnamefont
  {Moon}}\ and\ \bibinfo {author} {\bibfnamefont {M.}~\bibnamefont {Koshino}},\
  }\bibfield  {title} {\bibinfo {title} {Optical absorption in twisted bilayer
  graphene},\ }\href {https://doi.org/10.1103/PhysRevB.87.205404} {\bibfield
  {journal} {\bibinfo  {journal} {Phys. Rev. B}\ }\textbf {\bibinfo {volume}
  {87}},\ \bibinfo {pages} {205404} (\bibinfo {year} {2013})}\BibitemShut
  {NoStop}%
\bibitem [{Note2()}]{Note2}%
  \BibitemOpen
  \bibinfo {note} {$t_\perp $ is chosen so that for $\lambda _\perp =1$ the
  coupling for interlayer vertical hoppings has the same numerical value as in
  \cite {PhysRevB.85.195458, PhysRevB.87.205404}}\BibitemShut {NoStop}%
\bibitem [{\citenamefont {Moldovan}\ \emph {et~al.}(2020)\citenamefont
  {Moldovan}, \citenamefont {Andelkovi\'{c}},\ and\ \citenamefont
  {Peeters}}]{dean_moldovan_2020_4010216}%
  \BibitemOpen
  \bibfield  {author} {\bibinfo {author} {\bibfnamefont {D.}~\bibnamefont
  {Moldovan}}, \bibinfo {author} {\bibfnamefont {M.}~\bibnamefont
  {Andelkovi\'{c}}},\ and\ \bibinfo {author} {\bibfnamefont {F.}~\bibnamefont
  {Peeters}},\ }\href {https://doi.org/10.5281/zenodo.4010216} {\bibinfo
  {title} {{pybinding v0.9.5: a Python package for tight- binding
  calculations}}} (\bibinfo {year} {2020}),\ \bibinfo {note} {{This work was
  supported by the Flemish Science Foundation (FWO-Vl) and the Methusalem
  Funding of the Flemish Government.}}\BibitemShut {Stop}%
\bibitem [{Note3()}]{Note3}%
  \BibitemOpen
  \bibinfo {note} {As far as $\lambda _H$ is concerned, the request is to be in
  the Chern insulator regime. In the absence of an on-site chemical potential,
  this condition is met, in the full Haldane model, for $\lambda _{H}/t \in
  ]0,1/3[$. However, a sizable $\lambda _H>0.1t$ is beneficial for making the
  interpretation clearer, since for smaller values the topological edge states
  in the zigzag and bearded edges become reminiscent of the non-dispersive ones
  that characterize the $\lambda _H=0$ regime. Moreover, for our sample sizes
  we have explicitly checked the stability of the results for $\lambda _\perp
  \in [1,3]$. Bigger sample sizes allow for a larger parameter range, since the
  hybridization of the corner modes is reduced by their increased spatial
  separation.}\BibitemShut {Stop}%
\bibitem [{Note4()}]{Note4}%
  \BibitemOpen
  \bibinfo {note} {More precisely, the gaps would open around $E/t\approx +0.2$
  and $E/t\approx -0.25$. The tiny asymmetry in the gap openings is due to the
  interlayer hopping term which, once normalized with respect to $t$, has a
  negative coefficient ($t^\perp =-0.178 t$). This makes so that the whole
  spectrum is slightly shifted down. For simplicity, in what follows we will
  always denote the energy of the two gaps just as $E/t\approx \pm
  0.2$}\BibitemShut {NoStop}%
\bibitem [{Note5()}]{Note5}%
  \BibitemOpen
  \bibinfo {note} {Both in the tight binding calculation and in the low energy
  effective theory, a change of $\lambda _H$ renormalizes the Fermi velocity of
  the edge modes in the bearded and zigzag cases, hence slightly modifying the
  size of the energy gaps. In the zigzag-armchair case, the location of the gap
  is also slightly shifted.}\BibitemShut {Stop}%
\bibitem [{\citenamefont {Bistritzer}\ and\ \citenamefont
  {MacDonald}(2011)}]{Bistritzer12233}%
  \BibitemOpen
  \bibfield  {author} {\bibinfo {author} {\bibfnamefont {R.}~\bibnamefont
  {Bistritzer}}\ and\ \bibinfo {author} {\bibfnamefont {A.~H.}\ \bibnamefont
  {MacDonald}},\ }\bibfield  {title} {\bibinfo {title} {Moir{\'e} bands in
  twisted double-layer graphene},\ }\href
  {https://doi.org/10.1073/pnas.1108174108} {\bibfield  {journal} {\bibinfo
  {journal} {Proceedings of the National Academy of Sciences}\ }\textbf
  {\bibinfo {volume} {108}},\ \bibinfo {pages} {12233} (\bibinfo {year}
  {2011})}\BibitemShut {NoStop}%
\bibitem [{\citenamefont {Rost}\ \emph {et~al.}(2019)\citenamefont {Rost},
  \citenamefont {Gupta}, \citenamefont {Fleischmann}, \citenamefont
  {Weckbecker}, \citenamefont {Ray}, \citenamefont {Olivares}, \citenamefont
  {Vogl}, \citenamefont {Sharma}, \citenamefont {Pankratov},\ and\
  \citenamefont {Shallcross}}]{PhysRevB.100.035101}%
  \BibitemOpen
  \bibfield  {author} {\bibinfo {author} {\bibfnamefont {F.}~\bibnamefont
  {Rost}}, \bibinfo {author} {\bibfnamefont {R.}~\bibnamefont {Gupta}},
  \bibinfo {author} {\bibfnamefont {M.}~\bibnamefont {Fleischmann}}, \bibinfo
  {author} {\bibfnamefont {D.}~\bibnamefont {Weckbecker}}, \bibinfo {author}
  {\bibfnamefont {N.}~\bibnamefont {Ray}}, \bibinfo {author} {\bibfnamefont
  {J.}~\bibnamefont {Olivares}}, \bibinfo {author} {\bibfnamefont
  {M.}~\bibnamefont {Vogl}}, \bibinfo {author} {\bibfnamefont {S.}~\bibnamefont
  {Sharma}}, \bibinfo {author} {\bibfnamefont {O.}~\bibnamefont {Pankratov}},\
  and\ \bibinfo {author} {\bibfnamefont {S.}~\bibnamefont {Shallcross}},\
  }\bibfield  {title} {\bibinfo {title} {Nonperturbative theory of effective
  hamiltonians for deformations in two-dimensional materials: Moir\'e systems
  and dislocations},\ }\href {https://doi.org/10.1103/PhysRevB.100.035101}
  {\bibfield  {journal} {\bibinfo  {journal} {Phys. Rev. B}\ }\textbf {\bibinfo
  {volume} {100}},\ \bibinfo {pages} {035101} (\bibinfo {year}
  {2019})}\BibitemShut {NoStop}%
\end{thebibliography}
\end{document}